\documentclass[sigconf]{acmart}

\AtBeginDocument{%
  \providecommand\BibTeX{{%
    \normalfont B\kern-0.5em{\scshape i\kern-0.25em b}\kern-0.8em\TeX}}}

\usepackage{amsmath,amsfonts}
\usepackage{algorithmic}
\usepackage{bmpsize}
\usepackage{xspace}
\usepackage{lipsum}
\usepackage{float}
\usepackage[subtle,tracking=normal]{savetrees}
\usepackage{tikz}
\usepackage[english]{babel}
\usepackage{subcaption}
\usepackage{makecell}
\usepackage[ruled,vlined,linesnumbered]{algorithm2e}
\usepackage{multirow}
\usepackage{balance}
\usepackage{xcolor}
\usepackage{graphicx}
\usepackage{hyperref}
\usepackage[frozencache,cachedir=minted_cache]{minted}

\hypersetup{                    
  colorlinks,
  linkcolor={red!70!black},
  citecolor={red!70!black},
  urlcolor={red!70!black}
}

\newcommand{\pprobe}[0]{Prime-and-Probe}
\newcommand{\figtext}[2]{Fig.~\hyperref[#1]{\getrefnumber{#1}#2}}

\newcommand{\avgstd}[2]{$#1$ ($\sigma =\xspace#2$)}
\newcommand{\pt}[1]{$#1$\%}
\newcommand{\codeline}[1]{\f{\ref{#1}}}
\newcommand{\libgcryptversion}[0]{\f{1.7.0}\xspace}

\newcommand{\boldhline}[0]{\Xhline{2\arrayrulewidth}}

\def\f#1{{\texttt{#1}}}

\def\s{\f{s}}

\newcommand{\dejavu}[0]{D\'{e}j\`{a} Vu\xspace}

\newcommand{\TSX}[0]{Our-Prime+Abort\xspace}
\newcommand{\PageFault}[0]{Our-PF\xspace}
\newcommand{\PageCache}[0]{Our-PFCa\xspace}
\newcommand{\TSXCache}[0]{Our-Prime+AbortCa\xspace}
\newcommand{\CacheOnly}[0]{Our-Ca\xspace}

\newcommand{\CacheAttack}[0]{Cache attack\xspace}
\newcommand{\PageAttack}[0]{Page-faults attack\xspace}
\newcommand{\TSXAttack}[0]{Prime+Abort\xspace}

\newcommand{\ideal}{Monitor++}

\usepackage{booktabs}

\begin{document}
	 \settopmatter{printacmref=false}
	 \renewcommand\footnotetextcopyrightpermission[1]{}
	
	\pagestyle{plain}

\title{On the Challenges of Detecting Side-Channel Attacks in SGX}

\author{Jianyu Jiang}
\affiliation{%
  \institution{The University of Hong Kong}
  \city{Hong Kong}
  \country{China}}
\email{jyjiang@cs.hku.hk}

\author{Claudio Soriente}
\affiliation{%
  \institution{NEC Laboratories Europe}
  \city{Heidelberg}
  \country{Germany}}
\email{claudio.soriente@neclab.eu}

\author{Ghassan Karame}
\affiliation{%
  \institution{Ruhr-University Bochum}
  \city{Bochum}
  \country{Germany}}
\email{ghassan@karame.org}

\begin{abstract}
Existing tools to detect side-channel attacks on Intel SGX are grounded on the observation that attacks affect the performance of the victim application. As such, all detection tools monitor the potential victim and raise an alarm if the witnessed performance (in terms of runtime, enclave interruptions, cache misses, etc.) is out of the ordinary.

In this paper, we show that monitoring the performance of enclaves to detect side-channel attacks may not be effective. Our core intuition is that all monitoring tools are geared towards
an adversary that interferes with the victim's execution
in order to extract the most number of secret bits (e.g., the
entire secret) in one or few runs. They cannot, however, detect an adversary that leaks smaller portions of the secret---as small as a single bit---at each execution of the victim.  In particular, by minimizing the information leaked at each run, the impact of any side-channel attack on the application's performance is significantly lowered---ensuring that the detection tool does not detect an attack. By repeating the attack multiple times, each time on a different part of the secret, the adversary can recover the whole secret and remain undetected. Based on this intuition, we adapt known attacks leveraging page-tables and L3 cache to bypass existing detection mechanisms. We show experimentally how an attacker can successfully exfiltrate the secret key used in an enclave running various cryptographic routines of \f{libgcrypt}. Beyond cryptographic libraries, we also show how to compromise the predictions of enclaves running decision-tree routines of OpenCV. Our evaluation results suggest that performance-based detection tools do not deter side-channel attacks on SGX enclaves and that effective detection mechanisms are yet to be designed.
\end{abstract}

\maketitle

\section{Introduction} \label{sec:intro}

Intel Software Guard Extensions (SGX) enables applications to execute in isolation from other software on the same platform, including the OS. SGX-enabled processors run applications in so-called \emph{enclaves} and provide them with encrypted runtime memory, encrypted storage, and mechanisms to issue authenticated statements on the enclave software configuration. As such, a number of practitioners believe that Intel SGX is particularly suited for cloud deployments since it allows to outsource applications to the cloud, with the assurance that outsourced applications run untampered and their data is not available to any (privileged) software on the same host.

Previous work has, however, shown that Intel SGX exhibits a number side-channels that, when coupled with an adversary that controls the OS, allow for effective leakage of enclave secrets~\cite{moghimi17ches,schwarz17dimva,wang17ccs,brasser17woot,xu15sp}.
Alongside attacks, the research community has proposed a number of prevention~\cite{hyperrace:sp18,shinde16asiaccs,obfuscuro:ndss19,cloak:security17,drsgx:acsac19} and detection mechanisms~\cite{tsgx:ndss17,oleksenko18atc,chen17asiaccs}---the former having usually much higher overhead compared to the latter.
To the best of our knowledge, all detection mechanisms are grounded on the observation that side-channel attacks affect the performance of the victim application (e.g., by increasing the number of enclave interruptions) and, therefore, signal an attack when the witnessed performance is anomalous.

In this paper, we show that such detection tools may not be effective at detecting side-channel attacks on SGX enclaves. Namely, existing detection mechanisms are geared towards an adversary that interferes with the victim's execution in order to extract the most number of secret bits (e.g., the entire secret) in one or few runs. Such an attack strategy has a significant impact on the victim's performance, effectively allowing detection mechanisms to notice a deterioration in performance (e.g., in terms of runtime, enclave interruptions, cache misses, etc.) and signal an attack.

Our core intuition is that an adversary can leak smaller portions of the secret---as small as a single bit---at each execution of the victim, so as to minimize the impact on its performance and, therefore, remain undetected. More specifically, we show that an adversary can profile a victim enclave, thereby identifying the precise moment during the victim's execution when a specific part of the secret can be leaked via a side-channel attack. For example, if the victim runs the popular square-and-multiply algorithm, we show that the attacker can infer the moment when the $i$-th loop is being executed---i.e., when the $i$-th secret bit is being processed---and execute a side-channel attack at that time to leak the secret bit, without affecting the performance of the victim. By running the victim multiple times and leaking a different part of the secret at a time, our technique can recover the whole secret while remaining undetected.

Based on this intuition, we adapt known attacks leveraging page-tables, L3 cache, and a combination of the two, and evaluate their performance on routines of \f{libgcrypt} (namely, \f{mpi\_powm} and \f{mpi\_ec\_dup\_point}) used by popular cryptographic primitives such as ElGamal, RSA, and EdDSA. We also apply our attack strategy on non-cryptographic software and evaluate how to leak predictions of enclaves running decision-tree routines of OpenCV~\cite{opencv}.
Our results show that our strategy recovers up to \pt{100} of a secret key used in \f{libgcrypt} routines, depending on the type of side-channel exploited, and with marginal impact on the victim's performance (as low as one extra Asynchronous EXit (AEX) or roughly 40 cache misses per run). In case of a victim using the decision-tree routines of OpenCV  to predict handwritten digits of the MNIST data-set~\cite{mnist}, our attack strategy can correctly leak around \pt{55} of the predictions (whereas a ``standard'' side-channel attack, that is easily detected by available tools, reaches \pt{64} of leaked predictions).

We additionally show that an adversary using our attack strategy cannot be detected by existing detection tools such as T-SGX~\cite{tsgx:ndss17}, unless one tolerates a large number of false positives.
We also provide evidence that \emph{any} detection tool that monitors the performance of the victim is equally likely to fail. We do so by assuming a comprehensive tool (dubbed \ideal) that monitors all of the performance metrics proposed in literature and show that even such a tool cannot distinguish between a benign and a ``malicious'' execution.

Our results highlight that defenses that monitor performance metrics are not enough to detect side-channel attacks on Intel SGX enclaves. We therefore hope that our findings help avoiding additional (and probably unnecessary) cycles of defenses that monitor performance metrics and attacks that succeed at bypassing them.

The rest of this paper is organized as follows. In Section~\ref{sec:bg}, we overview necessary background information and related work on SGX. We describe the main intuition behind our attacks in Section~\ref{sec:attack} and we evaluate them against
\f{libgcrypt} and OpenCV in Sections~\ref{sec:impl}-\ref{sec:opencv}.
Finally, Section~\ref{sec:discuss} discusses possible defenses against our attacks and provides some concluding remarks.

\section{Background: Side-channel attacks on SGX} \label{sec:bg}

Previous research has shown that Intel SGX is vulnerable to side-channel attacks and that the Intel SGX threat model---by considering a malicious OS---allow for very effective attacks~\cite{brasser17woot,moghimi17ches,copycat:sec20,sgx-step}.

Proposed defenses work either as prevention or detection tools. Prevention techniques incur in high overhead~\cite{cloak:security17,obfuscuro:ndss19,shinde16asiaccs,drsgx:acsac19}, and sometimes can only prevent specific types of side-channels~\cite{hyperrace:sp18}.

Detection techniques have usually lower overhead and, to the best of our knowledge, they all use the same ``anomaly-based'' approach: they monitor the execution of the victim application and signal an attack in case of deviations from a ``normal'' execution.
Varys~\cite{oleksenko18atc} prevents L1/L2 cache-based attacks with core-reservation; at the same time, Varys detects attacks based on page-faults or interrupts by monitoring the number of AEXs so that an alarm is raised if their frequency is too high. Varys is currently part of a commercial product and its source-code is not available.
\dejavu~\cite{chen17asiaccs} detects attacks based on page-faults or interrupts by monitoring the execution time of the enclave.
\dejavu instruments the basic blocks of the enclave code to measure their execution time and an attack is ``detected'' if the total time deviates from the one of an execution in a benign environment. An incomplete version of \dejavu is available on github~\cite{dejavurepo}; we made contact with the authors to obtain the missing code, but they are no longer maintaining the project.
T-SGX~\cite{tsgx:ndss17} makes use of Transactional Synchronization eXtensions (TSX) to suppress page-faults notifications to the OS. When an interrupt or fault is thrown within a TSX transaction, T-SGX aborts and executes a user-defined handler. The handler of T-SGX keeps tracks of the number of aborts per transaction and raises an alarm if that number reaches a given threshold. The source code of T-SGX is available on github~\cite{tsgxrepo}.

\vspace {0.5 em}\noindent\textbf{Previous ``stealthy'' side-Channel attacks.} Previous work proposes side-channel attacks on enclaves that do not cause page-faults---thereby achieving stealthiness despite detection-tools that monitor page-faults. Jo Van et al.,~\cite{stealth:page:security17} monitor the \f{ACCESS} bit of the page-table to get the page access sequence of the victim without page-faults. As the \f{ACCESS} bit of a page-table is set only the first time the page is accessed (i.e., subsequent accesses do not modify the bit), the authors of~\cite{stealth:page:security17} force a TLB shootdown---by interrupting the enclave via inter-process-interrupts---to reset the \f{ACCESS} bit. The authors acknowledge that the number of interruptions during their attack is substantially higher than what is to be expected under
benign circumstances, and suggest that a detection tool may notice the attack by monitoring enclave interruptions rather than page-faults. Differently, the attack strategy we develop in this paper causes only a few interruptions of the victim and remains undetected.

Wang et al.,~\cite{wang17ccs} show that enclave interruptions can be minimized if TLB shootdown is achieved  by using a sibling hyperthread that probes memory addresses whose TLB entries are conflicted with the ones of the victim enclave. While the attack developed by~\cite{wang17ccs} cannot be detected by monitoring enclave interruptions, it requires the adversary and the victim to run on the same core. As such, the attack is not viable in case of detection tools that enforce core-reservation like Varys~\cite{oleksenko18atc} or \dejavu~\cite{chen17asiaccs}. Differently, our attack strategy does not require the adversary to run on the same core of the victim.

Another stealthy attack is Prime+Abort~\cite{prime:abort:sec17}.
Here, the idea is to use TSX as a ``watchdog'' so that whenever the victim touches a specific cache-line, the adversary receives an immediate hardware callback in the form of a transactional abort. In principle, the attack could be used to bypass detection mechanism that monitor asynchronous exits~\cite{oleksenko18atc,chen17asiaccs}  or that use TSX to suppress page-faults notifications to the OS~\cite{tsgx:ndss17}. However, a Prime+Abort attack could be easily spotted by monitoring cache misses.
In Section~\ref{sec:eval}, we show that a trivial modification to T-SGX~\cite{tsgx:ndss17} allows a victim to detect Prime+Probe attacks.

\section{Stealthy side-channel attacks} \label{sec:attack}

\subsection{Threat model} \label{sec:attack:threat}
We assume that the adversary has the victim code available (e.g., the code belongs to a library or an open-source implementation), controls the OS where the enclave is running, and can execute the victim enclave arbitrarily many times. Such assumptions are similar to the ones found in related work~\cite{wang17ccs,xu15sp,moghimi17ches} and capture a realistic cloud deployment where an application is uploaded by its owner to the cloud provider, and part of the application code (e.g., a decryption routine) runs in an enclave. After attestation and secret provisioning by the application owner, the cloud provider can (re-)start the application or trigger the routine running in the enclave arbitrarily many times.

\subsection{Main intuition}

Detection tools for (known) side-channel attacks build on the intuition that attacks are likely to alter the performance of the victim application. As a consequence, almost all detection tools monitor the performance of the potential victim, and signal an attack if the witnessed performance is anomalous.

We show that this intuition is not accurate. More precisely, we show that an adversary can bypass these tools while minimizing the effect on the victim's performance by ``spreading'' the attack across multiple runs. This can be done when the adversary extracts specific portions of the secret, as small as a single bit, at each run of the victim enclave. By minimizing the information leaked at each run, the impact of the attack on the victim's performance is also lessened---so that the detection tool notices no performance anomaly. This strategy is repeated for a number of times---each time leaking a different portion of the secret---to eventually recover the full secret.

In particular, we denote the enclave secret by $\s=\s_1,\dots,\s_n$, where each $\s_i$ could be a single bit or multiple ones. Moreover, assume the victim code is split into $n$ segments $S_1,\dots,S_n$, such that segment $S_i$ processes $\s_i$. Here, the application is  executed $n$ times. During the $i$-th run, the attacker launches a side-channel attack while the victim is executing segment $S_i$, in order to leak $\s_i$. As the attack only runs for a small time-window, the victim's performance is only marginally affected.

Developing the aforementioned strategy entails a number of challenges and requires the adversary to mount a side-channel attack only during the time-window when the victim is executing code segment $S_i$. One option would be to precisely control the victim's execution by using single-stepping frameworks like SGX-Step~\cite{sgx-step}. However, side-stepping the victim generates a large number of page-faults---allowing a tool that monitors the number of AEXs to detect such an attack. To remedy this, we take a different approach and design an offline automated profiling phase to learn the time-interval when the victim is executing a specific code segment $S_i$. In what follows, we detail the offline profiling phase and the design choices we made to minimize errors.

\subsection{Application profiling} \label{sec:attack:window}
Let $T_i$ be the time when the victim starts code segment $S_i$. Note that a segment is a logical execution unit and different segments may execute the same code, but on different portions of the secret. For example, in the square-and-multiply routine, each segment corresponds to one execution of the main loop and processes one secret bit.

In an ideal scenario, the execution time of each code segment is constant, i.e., $T_{i+1} - T_{i} = c$. Thus, segment $S_i$ starts at time $T_i = (i - 1)\cdot c$, for some constant $c$. More generally, the execution time of a code segment may depend on the code itself, as well as the portion of the secret it processes. Thus, we model the execution time of segment $S_i$ as a function $t_i(\s_i)$, and set the start time of segment $S_i$ as $T_i=\sum_{j<i}t_j(\s_j)$.

As an example, Figure~\ref{algo:segment} shows a simple code segment with a conditional branch on the $i$-th bit of variable \f{secret} and three different function calls (\f{m}, \f{g}, and \f{k}). If functions \f{m}, \f{g} and \f{k} have no conditional branches nor loops, we can use constants $c_\f{m}$, $c_\f{g}$, and $c_\f{k}$, to model their execution time. Thus, $t_i(\s_i)=c_\f{m} + \s_i \cdot c_\f{g} + (1 - \s_i) \cdot c_\f{k}$. In case any of the functions \f{m}, \f{g}, \f{k} has a loop or a conditional branch, we would recursively profile its execution time in a similar fashion.

Once we have the function $t_i(\s_i)$ that models the execution of $S_i$, we assess its values by running $S_i$ multiple times, and by using a different assignment of $\s_i$ each time. For example, to measure the execution time of the code in Figure~\ref{algo:segment}, we run the segment twice: once with $\s_i=0$ and once with $\s_i=1$. (We actually use multiple runs with the same configuration of variables, in order to make our measurements more robust.) Note that the enumeration of all possible configurations of the part of the secret processed by a code segment is feasible because each segment is likely to process only one or a few secret bits.
\smallskip

\noindent\textbf{Measuring execution time.} Since SGX cannot read the time-stamp counter via \f{rdstc}, we cannot directly inject time measurement instructions into code segments. Furthermore, when measuring the execution time of each code segment we cannot interrupt the enclave, as context switches between enclave and non-enclave code incurs extra overhead compared to context switches between regular processes~\cite{sgx:clock,sgx:eval}.
We, therefore, create a logical clock by means of a timer thread. We inject instructions at the start and end of each segment, to set a binary variable at a memory address \f{Addr} outside of the enclave memory. A separate timer thread continually gets the system timestamp using \f{rdstc} and checks the value of the variable at \f{Addr}. If the variable is set to \f{1}, the timer thread remembers the current timestamp and reset \f{Addr} to \f{0}. By measuring the time interval between two reads of \f{Addr} that returned \f{1}, we can infer the time required to run one code segment. 
\smallskip

\noindent\textbf{Stabilizing execution time.} Running time of arbitrary code on general-purpose machines is far from deterministic due to other software running on the same host.  Similar to~\cite{wang17ccs}, we reduce the noise due to other software on the same host by reserving a core for the victim enclave. Specifically, we use the \f{isolcpus} as boot-up option in Ubuntu. As a result, no tasks are assigned to the reserved core, nor it is interrupted for handling I/O. Furthermore, processes can be explicitly assigned to such cores (e.g., using \f{sched\_setaffinity}) and they can be interrupted by inter-processor interrupts. We also note that some detection tools~\cite{chen17asiaccs,oleksenko18atc} ensure core reservation to avoid side-channel attacks based on L1/L2 caches.

\begin{figure}[t]
\begin{minted}{c}
  void compute_on_s(char[] p, unsigned int secret) {  
    int tmp = m(p);
    for (int i = 0;i < NBITS;i++) {
      if ((secret & (1 << i)) != 0)
        g(tmp);
      else
        k(tmp);
    }
  }
\end{minted}
    \caption{Sample code segment.}
    \label{algo:segment}
    \vspace{-1 em}
\end{figure}

To reduce the noise due to state of the cache when the victim starts, we flush all caches before each execution. Although techniques such as speculative execution may still create differences in the state of the caches across different executions of the enclave, we have empirically verified that each run experiences almost the same amount of cache misses. We also disable dynamic frequency scaling and fix the CPU frequency to stabilize execution time.

\begin{figure}[t]
    \centering
    \includegraphics[width=\linewidth]{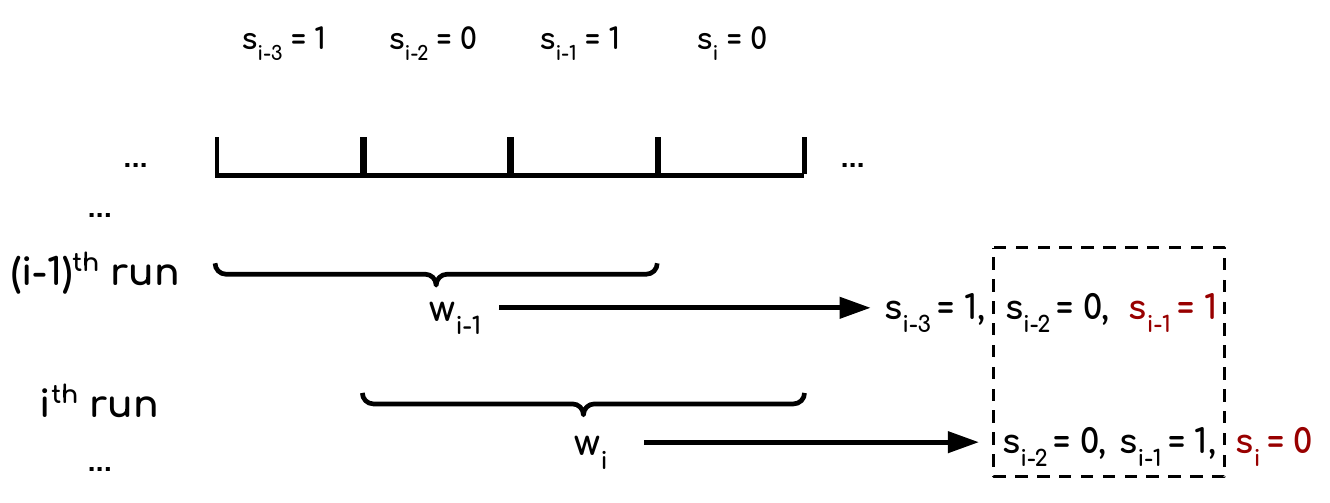}
    \vspace{-2 em}
    \caption{Alignment using a sliding window of size $w=3$.}
    \vspace{-1 em}
    \label{fig:fcaw}
\end{figure}

By combining core-reservation with cache-flushing and a fixed CPU frequency, we manage to stabilize the execution time of the victim (i.e., within $0.1\%$).

\vspace{0.5 em}\noindent\textbf{Improving attack accuracy.} The accuracy of our technique relies on the correct estimation of $T_i$---when we start the side-channel attack to learn $\s_i$---and the correct guess of $\s_i$.

Clearly, an error when estimating $T_i$ leads to a mis-alignment between the attack and the victim that, in turn, leads to unpredictable errors in inferring the secret $\s_i$. An error when inferring $\s_i$ may lead to an error in the estimation of $T_j$ for $j>i$ since the start time of segment $S_j$ may depend on the value of $\s_1,\dots,\s_{j-1}$.

One possible option to avoid miss-alignments between the victim and the attack is to rely on deterministic signals thrown by the enclave such as page-faults, page \f{ACCESS} bit, TSX aborts and so on. For example, the attacker may invalidate a page that is required by the victim at the start of $S_i$ so that a page-fault is thrown when the victim starts executing that segment. Alternatively, alignment errors may be corrected by attacking multiple consecutive segments at a time by using a sliding window. This basic idea is shown in Figure~\ref{fig:fcaw}. Let $w_i$ be the window attacking segments $S_{i-w+1},\ldots,S_{i}$ so to obtain bits $\s_{i-w+1},\ldots,\s_{i}$ and, without loss of generality, assume the step of the window to be 1. Then, we compare the guess for bits $\s_{i-w+1},\ldots,\s_{i-1}$ obtained when attacking window $w_i$, with the guess for the same bits obtained when attacking window $w_{i-1}$ (i.e., when attacking segments $S_{i-w},\ldots,S_{i-1}$). If the two bit sequences match, then we assume that window $w_i$ is well aligned and treat the guess for the last bit of the window (i.e., $\s_i$) as valid; otherwise we assume $w_i$ is not aligned with the victim and discard the guess for $\s_i$.

Note that attacking larger windows may have an impact on the victim's performance that could allow a detection mechanism to spot the attack. Also, our attack with $w = n$ becomes similar to a ``standard'' side-channel attack that tries to leak all secret bits at once.

In order to improve the accuracy of our technique, we can also increase the number of times we attack a given segment. That is, we run the victim $k$ times and run the attack on the same segment $S_i$ (or segment window $w_i$). We therefore obtain several samples for $\s_i$ and use heuristics to improve the accuracy of our guess.
\smallskip

\noindent\textbf{Automatic profiling.} To automate the profiling process, we expose two macros, \f{SEGMENT\_START(secret)} and \f{SEGMENT\_END}, for annotating the start and end of one segment, along with the portion of the secret consumed by that segment. These macros are then compiled with the victim code to generate a corresponding time-measuring code that records the execution time. During the profiling process, the portion of the secret used by a segment---usually one or a few bits---is enumerated and fed to the time-measuring code. For each configuration of the secret value, the time-measuring code generates a report with the execution time. The profiling process repeats to collect sufficient reports for stably modeling execution time of the victim.

\section{Compromising secrets in \f{libgcrypt}}
\label{sec:impl}

We now show how to instantiate the strategy described earlier on cryptographic routines of \f{libgcrypt}, namely \f{mpi\_powm} (used in ElGamal, RSA, and DSA) and \f{mpi\_ec\_mul\_point} (used in EdDSA). We leverage a side-channel based on time~\cite{llcattack:sp15,wang17ccs}, one based on memory access pattern~\cite{shinde16asiaccs,stealth:page:security17}, and a combination of the two. In what follows, we use \f{libgcrypt} version \f{1.7.0}; the side-channels we exploit are present in \f{mpi\_powm} up to version \f{1.8.6}, and in \f{mpi\_ec\_mul\_point} up to version \f{1.7.5}. We stress however that the above side-channels are mere examples to showcase the effectiveness of our strategy; our techniques are independent of the underlying side-channel and could use any other workable side-channel.

\begin{figure}[t]
\begin{minted}{c}
  void _gcry_mpi_powm (gcry_mpi_t res, gcry_mpi_t base, 
                                  gcry_mpi_t expo, gcry_mpi_t mod) {
    /* ... */
    gcry_mpi_t e = expo;
    int esec  = mpi_is_sec(expo);
    for (;e != 0;e = (e << 1)) {      @\label{elgamal:loop:start}@
      _gcry_mpih_sqr_n_basecase( xp, rp, rsize ); @\label{elgamal:base}@
      if(esec || (mpi_limb_signed_t)e < 0 ) { @\label{elgamal:branch}@
        /*mpihelp_mul( xp, rp, rsize, bp, bsize );*/ @\label{elgamal:branch:start}@
        if( bsize < KARATSUBA_THRESHOLD ) {
          _gcry_mpih_mul ( xp, rp, rsize, bp, bsize );
        } else {
          _gcry_mpih_mul_karatsuba_case (xp, rp, rsize, 
                                                          bp, bsize, 
                                                          &karactx);
        }
        xsize = rsize + bsize;
        if ( xsize > msize ) {
          _gcry_mpih_divrem(xp + msize, 0, xp, 
                                      xsize, mp, msize);
          xsize = msize;
        }
        if ( (mpi_limb_signed_t)e < 0 ) {
          tp = rp; rp = xp; xp = tp;
          rsize = xsize;
        }
      } @\label{elgamal:branch:end}@	
    } @\label{elgamal:loop:end}@
  }
\end{minted}
    \caption{\f{mpi\_powm} used in ElGamal, RSA and DSA.}
    \label{code:elgamal}
     \vspace{-1.5 em}
\end{figure}

\subsection{Side-channels of \textnormal{\f{mpi\_powm}}}
Figure~\ref{code:elgamal} shows the code of \f{mpi\_powm}. The routine has two side-channels, one based on time and another based on memory access pattern.

The loop (line \codeline{elgamal:loop:start} $\sim$ \codeline{elgamal:loop:end}) consumes one bit of the secret exponent per iteration and executes an extra computation (line \codeline{elgamal:branch:start} $\sim$ \codeline{elgamal:branch:end}) if that bit is \f{1} (line \codeline{elgamal:branch}). Thus, an adversary can infer the secret bit of the exponent being processed, by inferring the time to complete one loop iteration. Note that if \f{esec} is \f{1}, then the exponent is stored in secure memory, and the conditional branch is always executed to eliminate side-channels. However, if \f{xvalue} is provided as input by the user (e.g., when the key-pair is generated from a passphrase), then \f{libgcrypt} does not store the exponent in secure memory so that side-channels are not eliminated.

Alternatively, the secret bit of the exponent can be leaked by monitoring access to memory pages that store the code required by the \f{if}-branch of the routine. Let \f{A}, \f{B}, \f{C} be the addresses of \f{mpi\_powm}, \f{mpi\_mpih\_sqrt\_n\_basecase} and \f{mpihelp\_mul}, respectively. One iteration of the loop where the exponent bit is \f{1}, shows a memory access sequence like \f{ABCAC|AB}, whereas if the exponent bit is \f{0}, the observed memory access sequence is like \f{ABC|AB}. In these examples, memory accesses after \f{|} belong to the next iteration of the loop. Also, note that \f{mpi\_mpih\_sqrt\_n\_basecase} calls \f{mpihelp\_mul}, so there will always be an access to address \f{C} after \f{B}. One could infer the memory access sequence either by observing page-faults or cache accesses.

\begin{figure}[tb]
    \centering
    \includegraphics[width=.4\textwidth]{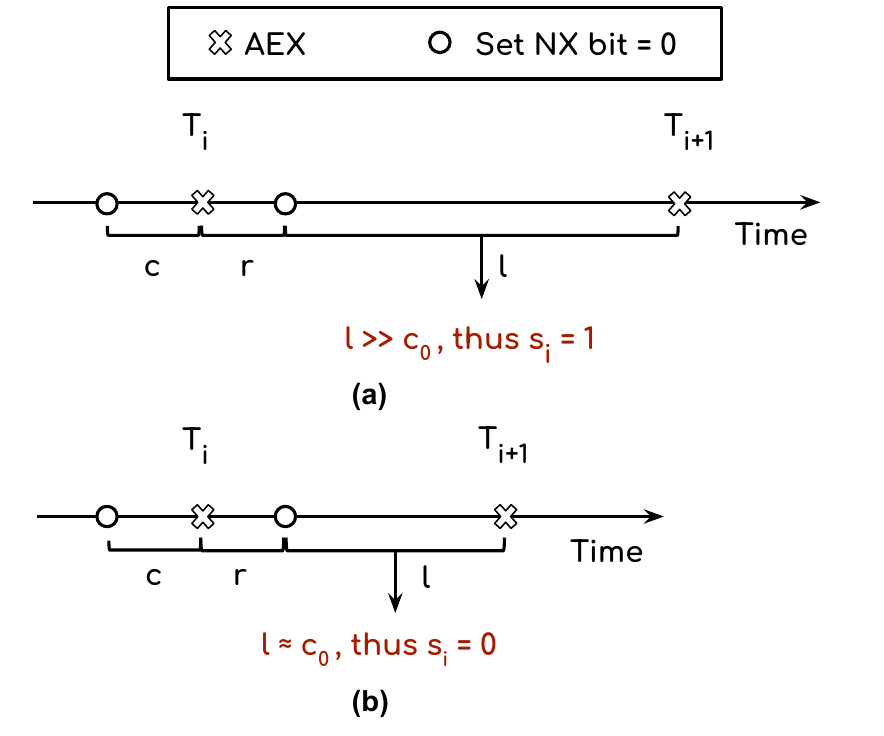}
    \caption{Workflow of \PageFault for (a) $\s_i=\f{1}$ or (b) $\s_i=\f{0}$.}
    \label{fig:fpf:flow}
    \vspace{-1.0 em}
\end{figure}

\subsection{Profiling of \textnormal{\f{mpi\_powm}}}

In order to profile \f{mpi\_powm}, we define each iteration of the main loop as one segment. Let $\s_i$ be the $i$-th exponent bit consumed in segment $S_i$. One iteration of the loop in \f{mpi\_powm} computes on \f{xp}, \f{rp} and $\s_i$. We found that all branches and loops in\\ \f{mpih\_sqr\_n\_basecase} have negligible impact on execution time, so we consider its runtime constant and we denote it by $c_{base}$. Thus, runtime of $S_i$ with $\s_i = 0$ is simply $c_{base}$. If $\s_i = 1$, the code executed (lines \codeline{elgamal:branch:start} $\sim$ \codeline{elgamal:branch:end}) has two branches. The first one is a conditional branch that, depending on the value of \f{bsize}, may run either \f{mpihelp\_mul} or \f{mpihelp\_mul\_karatsuba\_case}. We found that both paths take the same time so we model this time as a constant $c_{mpihelp}$. The second branch depends on \f{xsize} and \f{msize}. However, we found that the time taken to run \f{mpihelp\_divrem} is negligible, so we just ignore it. In a nutshell, the time to run segment $S_i$ is $t_i(\s)=c_{base} + \s_i \cdot c_{mpihelp}$ and $T_i = (i-1) \cdot c_{base} + \sum_{j<i} \s_j \cdot c_{mpihelp}$.

\subsection{Page-faults} \label{sec:fpf}

We start by describing an instantiation of our attack strategy that only uses page-faults and that leverages the timing side-channel of the victim; we denote this attack variant as \PageFault. More specifically, for $i=1,\dots,n$, we start the victim enclave running \f{mpi\_powm}, and use a single page-fault to stop it at the beginning of segment $S_i$. Next, we resume the victim and measure---again, using one page-fault---the time it takes to complete that segment, in order to learn the secret bit $\s_i$.

Figure~\ref{fig:fpf:flow} shows how the \PageFault attack strategy works. Assume the time it takes to run one iteration of the loop with exponent bit \f{0} and \f{1} is $c_0$ and $c_1$ (with $c_0<c_1$), respectively. We start the victim and set the \f{NX} bit of the page containing \f{mpih\_sqr\_n\_basecase}, right before time $T_i$ (i.e., at $T_i-c$, for some small constant $c$). As a result, the victim stops and throws a page-fault at the beginning of segment $S_i$---i.e., at the beginning of the $i$-th iteration of the loop. At this time, we resume the victim enclave, and sets again the bit \f{NX} of the page of \f{mpih\_sqr\_n\_basecase}. Therefore, the next page-fault will be thrown when the victim moves to the next segment. Hence, the time between the two page-faults is compared against $c_0$ and $c_1$, to decide the value of bit $\s_i$. Once we learn $\s_i$, we compute $T_{i+1}$ accordingly and move on to attack the next segment. This process is repeated for $i=1,\ldots, n$ in order to recover the whole secret. In practice, we also make sure that the \f{NX} bit is not set while the victim is running \f{mpih\_sqr\_n\_basecase}. We do so by ensuring that the bit is set $r$ ticks after the page-fault, where $r$ is the number of ticks required to run \f{mpih\_sqr\_n\_basecase}.

\subsection{Page-fault and cache} \label{sec:fpc}
\begin{figure}[t]
    \centering
    \includegraphics[width=.4\textwidth]{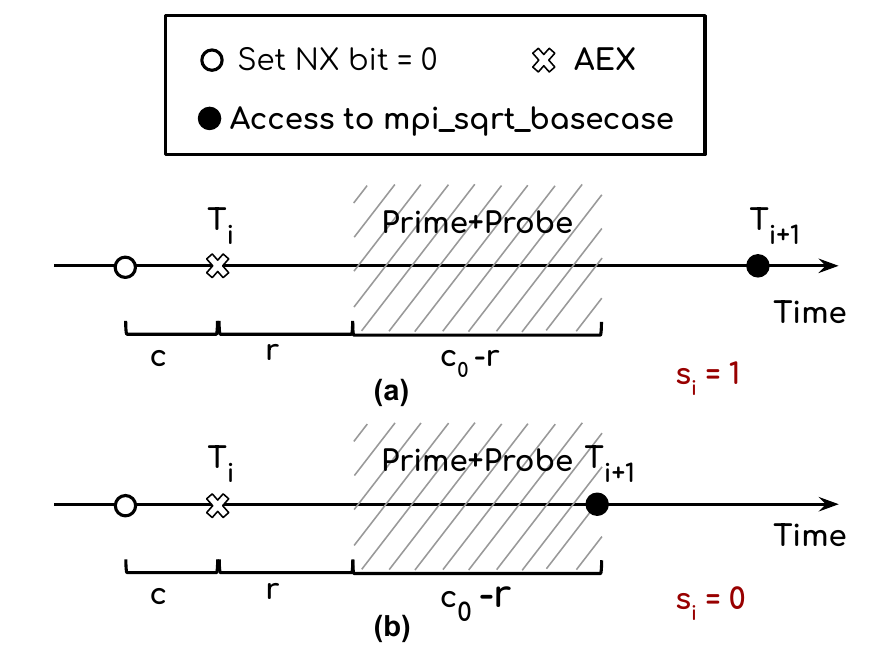}
    \caption{Workflow of \PageCache for (a) $\s_i=\f{1}$ or (b) $\s_i=\f{0}$.}
    \label{fig:fpc:flow}
    \vspace{-1.5 em}
\end{figure}

We now describe another attack variant, called \PageCache, that leverages both page-faults and cache misses. As \PageCache only leverages a side-channel based on memory access pattern, it could be used on routines that have no side-channel based on time.

Similar to \PageFault, we use one page-fault to stop the enclave at the beginning of a segment. Then, we use a \pprobe{} attack on L3 cache to infer the secret bit processed during the execution of that segment. We use L3 since most detection tools prevent L1/L2 attacks by occupying the entire core.

\begin{figure}[t]
    \center
    \includegraphics[width=.45\textwidth]{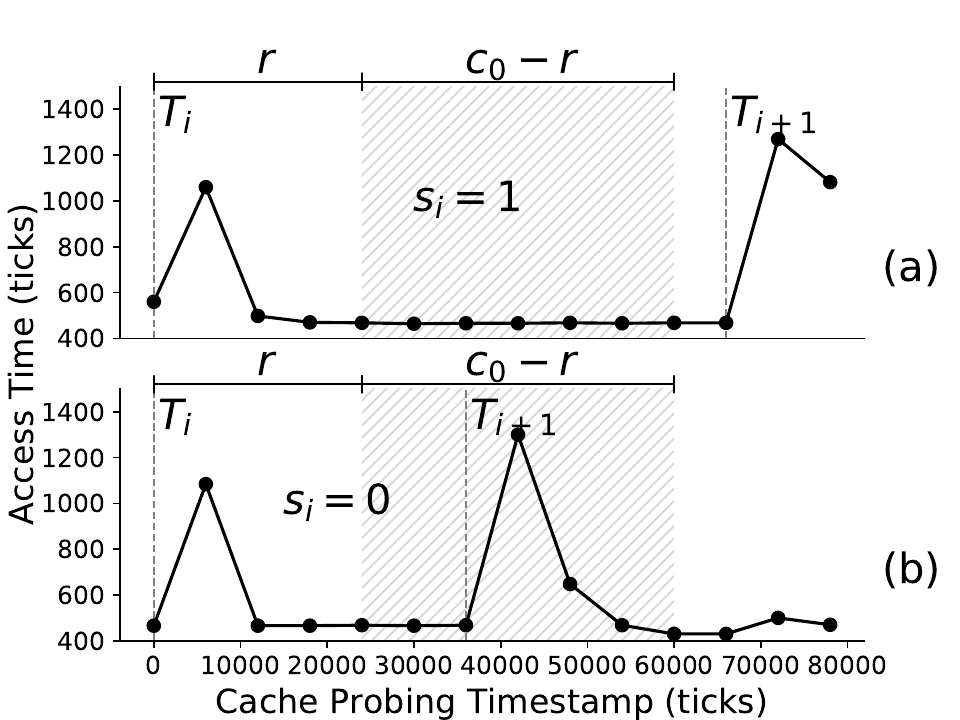}
    \caption{L3 probing pattern of \f{mpi\_powm} using \PageCache.}
    \label{fig:fpct:pattern}
        \vspace{-1.0 em}
\end{figure}

The workflow of \PageCache is shown in Figure~\ref{fig:fpc:flow}. Let $c_0$ and $c_1$ (with $c_0<c_1$) be the time it takes to run one loop of  \f{mpi\_powm} with secret bit \f{0} and \f{1}, respectively. We stop the enclave at $T_i-c$ by making the page of \f{mpih\_sqr\_n\_basecase} unavailable at that time; next, we resume the victim and wait for $r$ clock ticks to make sure that computation on \f{mpih\_sqr\_n\_basecase} is over. Now, the goal is to measure whether the next call to \f{mpih\_sqr\_n\_basecase} happens after time $c_0-r$ or $c_1-r$. To do so, we start a \pprobe{} attack on the address of \f{mpih\_sqr\_n\_basecase}, for a period of $c_0-r$. We construct the eviction set of the \pprobe{} using techniques from previous research~\cite{reverse:llc:raid15}. Figure~\ref{fig:fpct:pattern} shows the time to access the target cache set when the secret bit is \f{1} (a) or \f{0} (b). Here, it is clear that the victim has accessed the cache line of \f{mpih\_sqr\_n\_basecase} if the access time of the attacker to the eviction set is larger than $1000$ ticks. The first peak in each figure denotes the start of the i-th iteration, while the shaded area denotes the interval of $c_0-r$ ticks during which we run the \pprobe{} attack. Note that if $\s_i$ is \f{1}  (\figtext{fig:fpct:pattern}{a}) we do not witness any access to \f{mpih\_sqr\_n\_basecase} while running the \pprobe{} attack. In case $\s_i$ is \f{0} (\figtext{fig:fpct:pattern}{b}) we witness access to \f{mpih\_sqr\_n\_basecase} as the routine moves to the next iteration of the loop. Once we learn $\s_i$, we compute $T_{i+1}$ accordingly, and move on to attack the next segment.

\subsection{Cache-only} \label{sec:fca}

The attack strategies above use page-faults to temporally align the victim and attack threads. We now show how to run cache-only attacks on the victim enclave. In the sequel, we refer to this strategy as \CacheOnly.

Note that, using only cache to leak a specific portion of the victim's secret may be difficult because the adversary thread may not be aligned with the one of the victim; nevertheless, \CacheOnly is particularly effective with detection tools that monitor enclave exits (AEXs)~\cite{tsgx:ndss17,cosmix:atc19} as it enables the leakage of the secret without interrupting the victim at all.

\CacheOnly  works by starting a \pprobe{} attack on the address of \f{mpih\_sqr\_n\_basecase} right before $T_i$ and for $c_1$ ticks---the number ticks required to complete the loop iteration when the secret bit is \f{1}. If the attack thread experiences a peak in the time to access the target cache set, followed by a sufficient number of lows, we conclude that $\s_i=1$, whereas if the attack thread experiences two close peaks, we conclude that $\s_i=0$.

\begin{figure}[t]
    \center
    \includegraphics[width=.45\textwidth]{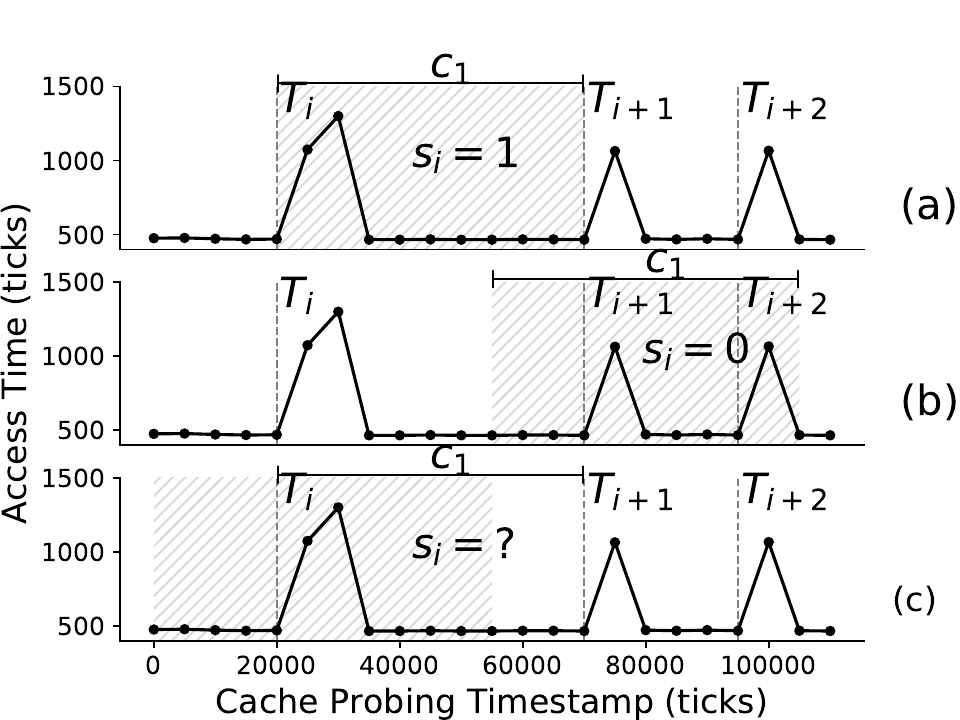}
    \caption{L3 probing pattern of \f{mpi\_powm} using \CacheOnly.}
    \label{fig:fca:pattern}
        \vspace{-1.0 em}
\end{figure}

A considerable challenge when using \CacheOnly lies in the fact that small errors when estimating $T_i$ leads to unpredictable cache patterns. This is shown in Figure~\ref{fig:fca:pattern}. In Figure~\ref{fig:fca:pattern}(a), the attack starts at the right time and the witnessed cache pattern does indeed support a correct guess of $\s_i$. Differently, in Figure~\ref{fig:fca:pattern}(b) the attack starts late and the adversary (mistakenly) estimates $\s_i$ to be \f{0}. Finally, in Figure~\ref{fig:fca:pattern}(c), the attack starts early, preventing the adversary from estimating the value of the secret bit.

We correct alignment errors between adversary and victim by using a sliding window technique as explained in Section~\ref{sec:attack}. That is, when attacking window $w_i$ (i.e., segments $S_{i-w+1},\ldots,S_{i}$) we start the \pprobe{} attack right before $T_{i-w+1}$ and we run it for $w c_1$ ticks---i.e. until the end of segment $S_i$. Next, we consider the estimate of $\s_i$ as valid only if the estimate of $\s_{i-w+1},\ldots,\s_{i-1}$ matches the estimate of the same bits when attacking window $w_{i-1}$. Finally, we also repeat the attack on the same window a number $k\geq 1$ of times in order to obtain multiple guesses for the same bit and use an heuristic to infer its actual value.

\subsection{Attacks on mpi\_ec\_mul\_point} \label{sec:eddsa}

We now briefly discuss how to adapt our attack strategy to the\\ \f{mpi\_ec\_mul\_point} routine of \f{libgcrypt} used in EdDSA. For each signature, this subroutine is used to compute scalar multiplication with a nonce that, if leaked, allows the recovery of the signing key. Note that our attack extracts one secret bit for each execution of the victim; hence, if the victim picks a fresh nonce at each execution, two bits extracted by our attack would be completely uncorrelated. Nevertheless, EdDSA is deterministic~\cite{eddsa} and the nonce is computed as function of the message to be signed and the signing key. Hence, by feeding a fixed message to the signing routine we ensure that the nonce is always the same and can extract one of its bits at each execution.

\definecolor{bgcolor}{rgb}{0.95,0.95,0.95}

Figure~\ref{code:eddsa} shows the code of \f{mpi\_ec\_mul\_point}. Note that the same routine is used to process both the nonce and the signing key (referred to as \f{scalar} in both cases). The leakage-free code (line \codeline{eddsa:secure:branch:start} $\sim$ \codeline{eddsa:secure:branch:end}) is used when processing the singing key, whereas the \f{else}-branch is taken to process the nonce. In the latter case, a secret-dependent branch (line \codeline{eddsa:dependent:branch}) can be abused to leak one bit of the (secret) nonce. Once the nonce and the corresponding signature are available, the signing key can be computed.

\noindent\textbf{Profiling \f{mpi\_ec\_mul\_point}.} Let segment $S_i$ be the i-th iteration of the loop. We found that there are no conditional loops nor branches in \f{gcry\_mpi\_ec\_dup\_point} that have noticeable impact on execution time, so we model its execution time with constant $c_{base}$. In case the bit of \f{scalar} being processed is \f{1}, the routine calls another constant-time function called \f{mpi\_ec\_add\_point} and we model its execution time with constant $c_{add}$. Therefore, the running time of the i-th loop iteration is $t_i(\s)=c_{base} + s_i \cdot c_{add}$, and the start time of the $i$-th segment is $T_i = (i-1) \cdot c_{base} + \sum_{j<i} s_j \cdot c_{add}$.

In practice, we must also accommodate for the first iteration of the loop that takes the \f{if}-branch; this iteration must fetch\\ \f{mpi\_ec\_add\_point} and its callees from the main memory and incurs in a time increase that we model with $c_{miss}$.
Thus the start time for the $i$-th segment becomes
$T_i = (i - 1) \cdot t_{base} + \sum_{j<i} s_j \cdot c_{add} + (\sum_{j=1}^{j=i-1} s_i \mod 2) \cdot t_{cache\_miss}$.
This extra time for the first loop that processes a \f{1} bit does not show in \f{mpi\_powm}, as the secret-dependent call to \f{mpi\_sqrt\_n\_basecase} is also called in other function before \f{mpi\_powm}.

When attacking \f{mpi\_ec\_mul\_point} we use the page\\ with \f{mpi\_ec\_dup\_point} to stop the enclave at the target segment. Further, we target \f{mpi\_ec\_dup\_point} when launching the \pprobe{} attack.

\begin{figure}[t]
\begin{minted}{c}
void _gcry_mpi_ec_mul_point ( mpi_point_t result, 
  gcry_mpi_t scalar, mpi_point_t point, mpi_ec_t ctx) {
  /* ... */
  if (mpi_is_secure (scalar)) { @\label{eddsa:secure:branch:start}@
    /* Oblivious Implementation */ 
  } else { @\label{eddsa:secure:branch:end}@
    /* Implementations with side-channels */ 
    for (j=nbits-1; j >= 0; j--) {
      _gcry_mpi_ec_dup_point (result, result, ctx); @\label{eddsa:duplicate:branch}@
      if (mpi_test_bit (scalar, j)) @\label{eddsa:dependent:branch}@
        _gcry_mpi_ec_add_points (result, result, point, 
                                 ctx);
    }
  }
}
\end{minted}
    \caption{\f{mpi\_ec\_mul\_point} used in EdDSA.}
    \label{code:eddsa}
     \vspace{-1 em}
\end{figure}

\section{Results  for \f{libgcrypt}} \label{sec:eval}

We instantiated the relevant routines of \f{libgcrypt} \libgcryptversion in SGX, by using Panoply~\cite{panoply:ndss17}. Our experiments were carried out with Ubuntu 18.04 on an Intel E3-1280 with $4$ physical cores and $32$GB RAM.

To assess the effectiveness of our attack strategy in evading existing detection tools, we compiled and run the two cryptographic routines using T-SGX with some engineering efforts. Since the other two detection tools available in literature are not released as open-source---Varys is part of a commercial product and \dejavu{} is no longer maintained---we also evaluate the effectiveness of our strategy on enclaves equipped with a comprehensive tool, dubbed \ideal, assumed to monitor all of the performance metrics proposed in literature (i.e., number of AEX, cache misses, and execution time). \ideal{} raises an alarm if the witnessed performance is anomalous. We note that cache misses are typically monitored via performance counters---a feature that is not currently available for SGX enclaves. Nevertheless, previous work has shown that non-SGX applications could use cache-misses to detect cache-based attacks~\cite{briongos18codaspy}; hence, we also include the number of cache misses among the performance metrics that are monitored by \ideal, to capture the possibility that it becomes available to future SGX applications.

\subsection{Profiling accuracy} \label{eval:stop:enclave}
We start by measuring the execution time of one loop of the victim routines---recall that a loop of \f{mpi\_powm} and \f{mpi\_ec\_mul\_point} is a code segment as defined in Section~\ref{sec:attack}. We do so by running each loop $100$ times with secret bit \f{0} and another $100$ times with secret bit \f{1}.

Our results show that, in case of using \ideal, a ``\f{0}-loop'' of \f{mpi\_powm} takes on average $46.4$k clock ticks ($\sigma=493.6$), while a ``\f{1}-loop'' takes on average $92.9$k clock ticks ($\sigma=122.2$). When instrumented with T-SGX, \f{mpi\_powm} takes slightly longer: $48.3$k clock ticks ($\sigma=530.1$) for a \f{0}-loop and $103.2$k clock ticks ($\sigma=251.3$) for a \f{1}-loop.

Routine \f{mpi\_ec\_mul\_point} with \ideal, takes on average $15.4$k clock ticks ($\sigma=378.1$) for a \f{0}-loop, and $39.2$k clock ticks ($\sigma=284.9$) for a \f{1}-loop. When instrumented with T-SGX, the function \f{mpi\_ec\_ec\_mul\_point} nearly double its computation time: it takes $38.8$k clock ticks ($\sigma=631.3$) and $92.0$k clock ticks ($\sigma=376.1$) for \f{0}-loop and \f{1}-loop, respectively.

Once we have the running times for \f{0}-loop and \f{1}-loop iterations, we validate the accuracy of our profiling technique by checking whether we can stop the enclave at the start of each loop. To do so, we fix a random $256$ bit secret and we execute the enclave $256$ times, each time stopping it at time $T_i-c$ (with $i=1,\dots,256$ and $c=5,000$ clock ticks).\footnote{Note that we can correctly estimate any $T_i$ because we know the value of the secret bits.} In order to learn the ground truth, we inject a counter into the code to keep track of the number of loop iterations thus far. This experiment is repeated $20$ times and we report the results in Table~\ref{tab:time:model}.

Stopping a victim enclave equipped with \ideal{} at the start of a loop, leverages the fact that the victim exposes page fault to the OS. However, if the victim uses T-SGX, we note that page-faults are dispatched to the enclave abort-handler so that the OS is not notified. We therefore leverage a technique similar to Prime+Abort~\cite{prime:abort:sec17}. In particular, we leverage TSX and setup a transaction with a cache set that conflicts with the memory address that starts the execution of a loop at the victim. As a result, our transaction aborts as soon as the victim starts a loop.

Our evaluation shows that stopping at a specific code segment an enclave running \f{mpi\_powm} with \ideal{} is more accurate (\pt{93.17}) than achieving the same if the enclave runs the routine instrumented with T-SGX (\pt{77.68}). This is because we may lose synchrony with the victim as T-SGX restarts a transaction. We observe the same behavior for \f{mpi\_ec\_mul\_point}: \pt{81.74} for the version using \ideal{} and \pt{67.33} for the version instrumented with T-SGX. A comparison between \f{mpi\_powm} and \f{mpi\_ec\_mul\_point} shows lower accuracy for the latter. This is because one loop of \f{mpi\_ec\_mul\_point} takes less time to complete compared to a loop of \f{mpi\_powm}---therefore, it is harder to hit the start of a specific loop iteration.
\begin{table}[t]
    \center
    \scalebox{1.05}{\begin{tabular}{l|c}
    \boldhline
                                & \textbf{Accuracy} \\
    \hline
    \f{mpi\_powm} (w/ \ideal)               & \avgstd{~\pt{93.17}}{~\pt{5.49}} \\
    \f{mpi\_powm} (w/ T-SGX)       & \avgstd{~\pt{77.68}}{~\pt{10.90}} \\
    \f{mpi\_ec\_mul\_point}  (w/ \ideal)   & \avgstd{~\pt{81.74}}{~\pt{4.52}}\\
    \f{mpi\_ec\_mul\_point} (w/ T-SGX)    & \avgstd{~\pt{67.33}}{~\pt{3.40}}\\
    \boldhline
    \end{tabular}}
    \caption{Accuracy when stopping the victim enclave at the beginning of a specific code segment.
    \label{tab:time:model}}
    \vspace{-2 em}
\end{table}

\begin{table*}[h]
    \center
     \scalebox{0.9}{
\begin{tabular}{c|c|c|c|c|c|c}

&  & \textbf{Detection Tool} &\textbf{Attack Accuracy}& \textbf{AEX/TSX Aborts}      & \textbf{L3 Cache-misses} & \textbf{Time (ms)} \\

\hline
\hline

\multirow{6}{*}{Our attacks}&\PageFault & \multirow{5}{*}{\ideal}                                     & \avgstd{\pt{85.5}}{~\pt{7.9}} & \avgstd{3.04}{0.20}   & \avgstd{125.91}{82.72}   &  \avgstd{5.67}{0.031}  \\

    &\PageCache&                                         & \avgstd{\pt{69.8}}{~\pt{6.1}} & \avgstd{2.70}{0.53}   & \avgstd{175.05}{135.05}     & \avgstd{5.64}{0.014} \\

    &\CacheOnly ($w = 3$)&                                 &\avgstd{\pt{76.3}}{~\pt{10.1}}& \avgstd{1.32}{0.46}   & \avgstd{164.05}{47.06}     & \avgstd{5.62}{0.010}  \\

    &\CacheOnly ($w = 5$)&                                 &\avgstd{\pt{89.14}}{~\pt{13.54}} & \avgstd{1.39}{0.48}    & \avgstd{218.81}{23.67}    & \avgstd{5.62}{0.0099} \\

    &\CacheOnly ($w = 9$)&        	                        &\avgstd{\pt{99.7}}{~\pt{0.5}}& \avgstd{1.30}{0.46}   & \avgstd{275.76}{38.02}   & \avgstd{5.62}{0.011}   \\
\cline{2-7}

    &\TSX & \multirow{1}{*}{T-SGX}  & \avgstd{\pt{71.03}}{~\pt{3.17}} & \avgstd{6.82}{10.38}   & \avgstd{431.76}{415.22}   &  \avgstd{5.71}{0.04}\\
	
\hline
\hline

\multirow{3}{*}{Standard attacks}& \PageAttack & \multirow{2}{*}{\ideal}   &\avgstd{\pt{97.9}}{~\pt{3.2}} & \avgstd{831.97}{99.69} & \avgstd{124.16}{21.78}   & \avgstd{11.50}{0.758} \\
    &\CacheAttack&                                         &\avgstd{\pt{89.5}}{~\pt{4.3}}& \avgstd{1.67}{0.71}       & \avgstd{2112.86}{82.10}   &  \avgstd{5.68}{0.0078} \\
\cline{2-7}
&\TSXAttack &\multirow{1}{*}{T-SGX}  & \avgstd{\pt{88.1}}{\pt{1.1}} & \avgstd{577.22}{99.43} & \avgstd{7834.45}{1984.30} & \avgstd{5.84}{0.09}\\

\hline
\hline

\multirow{6}{*}{No attack}&\f{mpi\_powm}&\multirow{3}{*}{\ideal} && \avgstd{2.441}{1.93}  & \avgstd{123.27}{82.91} & \avgstd{5.63}{0.017} \\
    &\f{mpi\_powm} (w/ GCC)&  &   & \avgstd{14.44}{10.97} & \avgstd{495.0}{290.11}    & \avgstd{5.78}{0.09} \\
    &\f{mpi\_powm} (w/ Redis)&                  && \avgstd{1.48}{0.70}                & \avgstd{6007.21}{510.83}                           & \avgstd{6.05}{0.48} \\

\cline{2-7}

    &\f{mpi\_powm}&\multirow{3}{*}{T-SGX}  && \avgstd{6.10}{21.76}                          & \avgstd{416.48}{410.76}                           & \avgstd{5.66}{0.02}  \\

    &\f{mpi\_powm} (w/ GCC)&  &   & \avgstd{188.67}{85.14} & \avgstd{394.14}{549.7}    & \avgstd{5.71}{0.60} \\

    &\f{mpi\_powm} (w/ Redis)&                   && \avgstd{121.25}{120.68}                  & \avgstd{16256.47}{5843.78}              & \avgstd{6.12}{0.25} \\

\hline
\hline

    \end{tabular}
     }
    \caption{Accuracy and performance metrics for \f{mpi\_powm}.}
    \vspace{-2em}
    \label{tab:elgamal:event}
\end{table*}

\begin{table*}[h]
    \center
    \scalebox{0.87}{
        \begin{tabular}{c|c|c|c|c|c|c}
    & &\textbf{Detection Tool} &\textbf{Attack Accuracy} &  \textbf{AEX / TSX Aborts}      & \textbf{L3 Cache-misses} & \textbf{Time (ms)} \\
\hline
\hline

\multirow{6}{*}{Our attacks}&\PageFault  &\multirow{5}{*}{\ideal} &\avgstd{\pt{69.6}}{\pt{3.3}}& \avgstd{3.01}{0.12}  & \avgstd{98.16}{14.12}        & \avgstd{6.31}{0.012}     \\

&\PageCache&       &\avgstd{\pt{64.4}}{\pt{2.7}}& \avgstd{2.33}{0.47}   & \avgstd{155.94}{112.04}       & \avgstd{6.30}{0.010}       \\

&\CacheOnly  ($w = 3$) &                               &\avgstd{\pt{86.4}}{\pt{12.91}}& \avgstd{1.60}{0.49}   & \avgstd{186.43}{16.99}        & \avgstd{6.29}{0.012}       \\

&\CacheOnly ($w = 5$)&                                 &\pt{100}& \avgstd{1.50}{0.49}   & \avgstd{201.41}{22.22}       & \avgstd{6.29}{0.011}      \\

& \CacheOnly ($w = 9$) &      	                        &\pt{100}& \avgstd{1.50}{0.50}   & \avgstd{249.38}{22.75}       & \avgstd{6.29}{0.012}        \\

\cline{2-7}

&\TSX & \multirow{1}{*}{T-SGX}&\avgstd{\pt{70.1}}{\pt{2.5}}& \avgstd{7.04}{9.25}  &\avgstd{219.07}{116.8}        & \avgstd{14.02}{0.48}        \\

\hline
\hline

\multirow{3}{*}{Standard attacks}
&\PageAttack&\multirow{2}{*}{\ideal} &\avgstd{\pt{99.6}}{\pt{0.22}}& \avgstd{499.28}{96.31}   & \avgstd{98.80}{21.09} & \avgstd{10.19}{0.76}  \\

&\CacheAttack&                      &\avgstd{\pt{96.8}}{\pt{5.0}}& \avgstd{1.47}{0.50}   & \avgstd{9684.87}{701.08}       & \avgstd{6.46}{0.011}          \\
\cline{2-7}
&\TSXAttack& \multirow{1}{*}{T-SGX} & \avgstd{\pt{98.9}}{\pt{1.8}} & \avgstd{695.89}{72.91} & \avgstd{18605.17}{2646.15} & \avgstd{13.53}{0.11} \\

\hline
\hline

\multirow{6}{*}{No attacks}&\f{mpi\_ec\_mul\_points}  &\multirow{3}{*}{\ideal}&   & \avgstd{2.71}{2.28}   & \avgstd{106.23}{60.79}       & \avgstd{6.29}{0.012}               \\

&\f{mpi\_ec\_mul\_points} (w/ GCC)&  &   & \avgstd{23.21}{27.83} & \avgstd{1246.31}{1331.89}    & \avgstd{6.30}{0.02} \\

&\f{mpi\_ec\_mul\_points} (w/ Redis)& &   & \avgstd{1.61}{0.83}   & \avgstd{6092.45}{1043.90}   & \avgstd{7.21}{1.02}          \\

\cline{2-7}
&\f{mpi\_ec\_mul\_points}  &\multirow{3}{*}{T-SGX}   &   & \avgstd{6.22}{34.48}   & \avgstd{158.93}{133.85}       & \avgstd{13.31}{0.29}          \\

&\f{mpi\_ec\_mul\_points} (w/ GCC)&  &   & \avgstd{377.93}{87.54} & \avgstd{817.54}{1331.89}    & \avgstd{13.67}{0.50} \\

&\f{mpi\_ec\_mul\_points} (w/ Redis)&       &   & \avgstd{226.91}{109.03}   & \avgstd{84941.38}{25664.85}       & \avgstd{17.69}{1.17}               \\

\hline
\hline

    \end{tabular}
    }
    \caption{Accuracy and performance metrics for \f{mpi\_ec\_mul\_point}.
    }
    \label{tab:eddsa:event}
    \vspace{-2 em}
\end{table*}

\subsection{Attack accuracy} \label{eval:attack:precision}

We evaluate the accuracy of our attack variants in recovering secret bits when the victim is either equipped with \ideal{} or with T-SGX. In case of T-SGX, we do not use attack variants that leverage cache since the original T-SGX paper does not address cache-based attacks~\cite{tsgx:ndss17}.

We fix a random $256$ bit secret and, for $i=1,\dots,256$, we recover the secret bit $\s_i$ by attacking the corresponding code segment $9$ times (i.e., for $9$ times we run the enclave and launch the side-channel attack from $T_{i}$ until $T_{i+1}$). Given the $9$ samples, we determine the secret bit based on majority voting. We repeat the experiment $10$ times and show the average accuracy and standard deviation in the column ``Attack Accuracy'' in Table~\ref{tab:elgamal:event} and Table~\ref{tab:eddsa:event} for \f{mpi\_powm}  and \f{mpi\_ec\_mul\_point}, respectively.
For comparison purposes, we also report the accuracy of ``standard'' side-channel attacks using either page-faults~\cite{wang17ccs} or L3 cache~\cite{wang17ccs,llcattack:sp15}. A standard attack refers to an attack that runs throughout the whole execution of the victim in order to recover the largest number of secrets bits in one execution. In case of standard attacks we also repeat the attack $9$ times and use majority voting to decide the value of each secret bit. Note that in case of routines instrumented with T-SGX, a standard page-faults attack does not work as T-SGX does not expose page-faults to the OS. In this case, we use the Prime+Abort attack of~\cite{prime:abort:sec17}.

\begin{figure*}[t]
    \centering
    \begin{subfigure}{.247\textwidth}
        \centering
        \includegraphics[width=\textwidth]{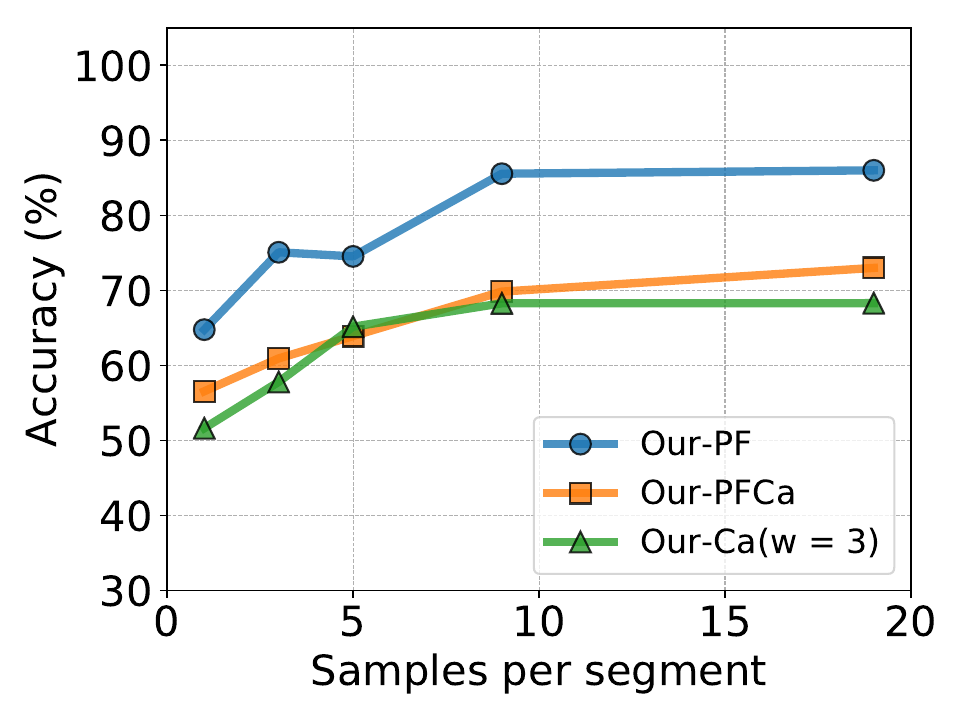}
        \caption{\f{mpi\_powm} \\ (w/ \ideal). \label{fig:elgamal:result:k}}
    \end{subfigure}%
    \begin{subfigure}{.247\textwidth}
        \centering
        \includegraphics[width=\textwidth]{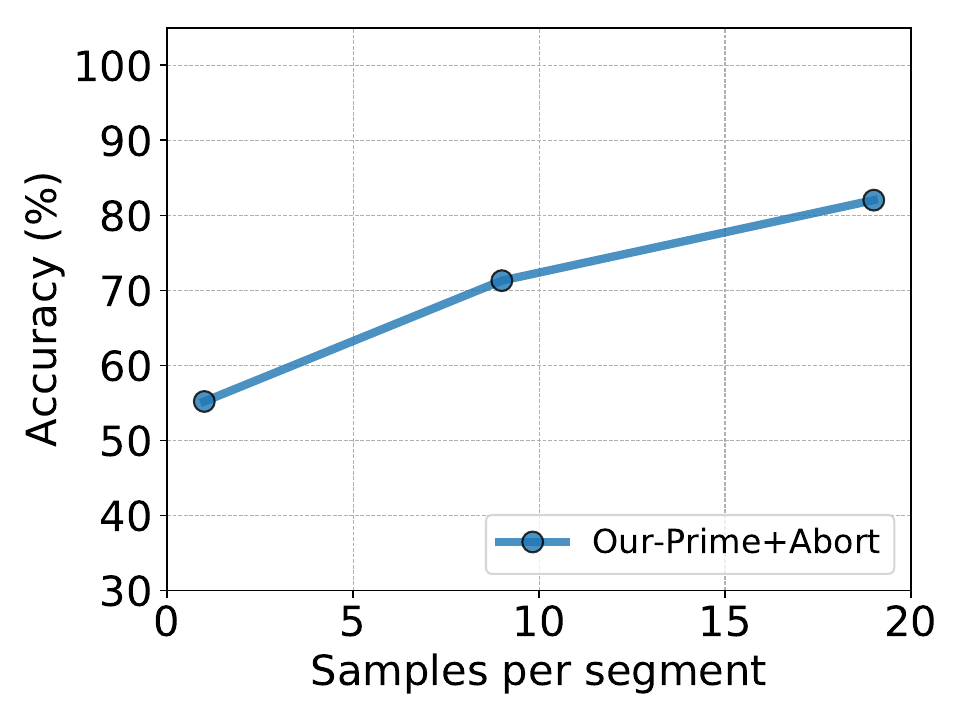}
        \caption{\f{mpi\_powm} \\ (w/ T-SGX). \label{fig:elgamal:result:k:T-SGX}}
    \end{subfigure}%
    \begin{subfigure}{0.247\textwidth}
        \centering
        \includegraphics[width=\textwidth]{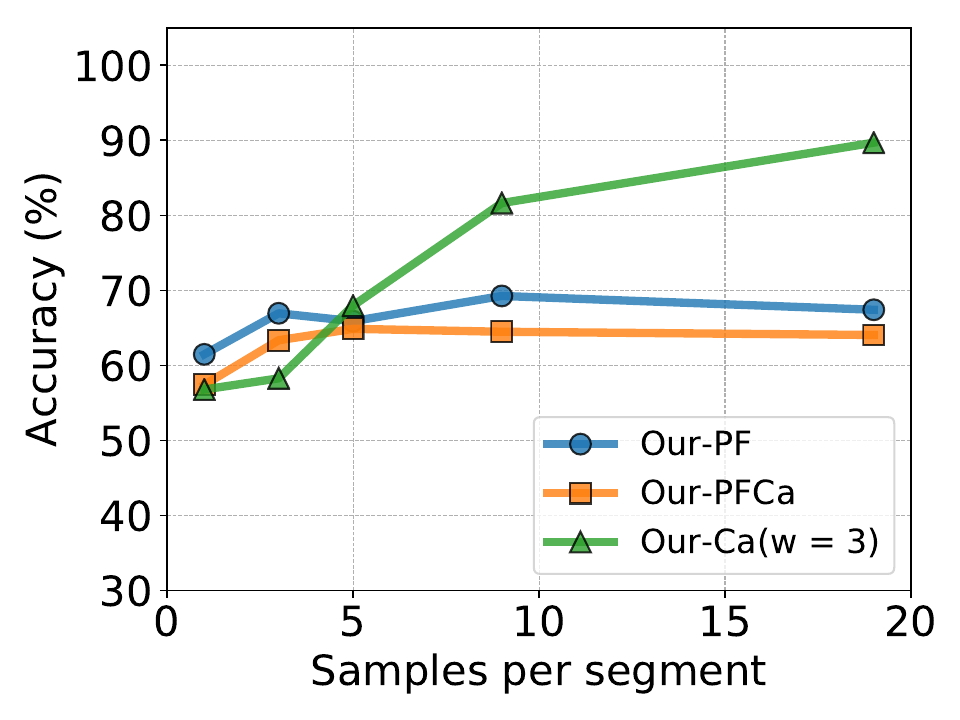}
        \caption{\f{mpi\_ec\_mul\_point} \\ (w/ \ideal).\label{fig:eddsa:result:k}}
    \end{subfigure}
    \begin{subfigure}{0.247\textwidth}
        \centering
        \includegraphics[width=\textwidth]{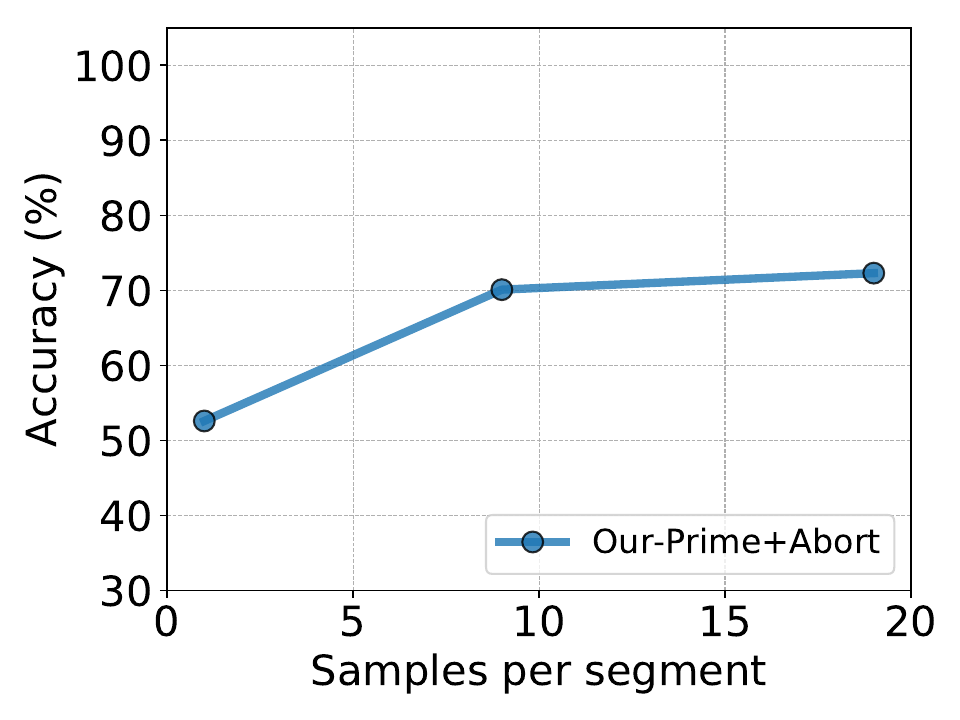}
        \caption{\f{mpi\_ec\_mul\_point} \\ (w/ T-SGX). \label{fig:eddsa:result:k:T-SGX}}
    \end{subfigure}
\caption{Accuracy versus number of samples per segment. \label{fig:k}}
\vspace{-1 em}
\end{figure*}

\begin{figure}[ht]
    \centering
    \begin{subfigure}{0.24\textwidth}
        \centering
        \includegraphics[width=\textwidth]{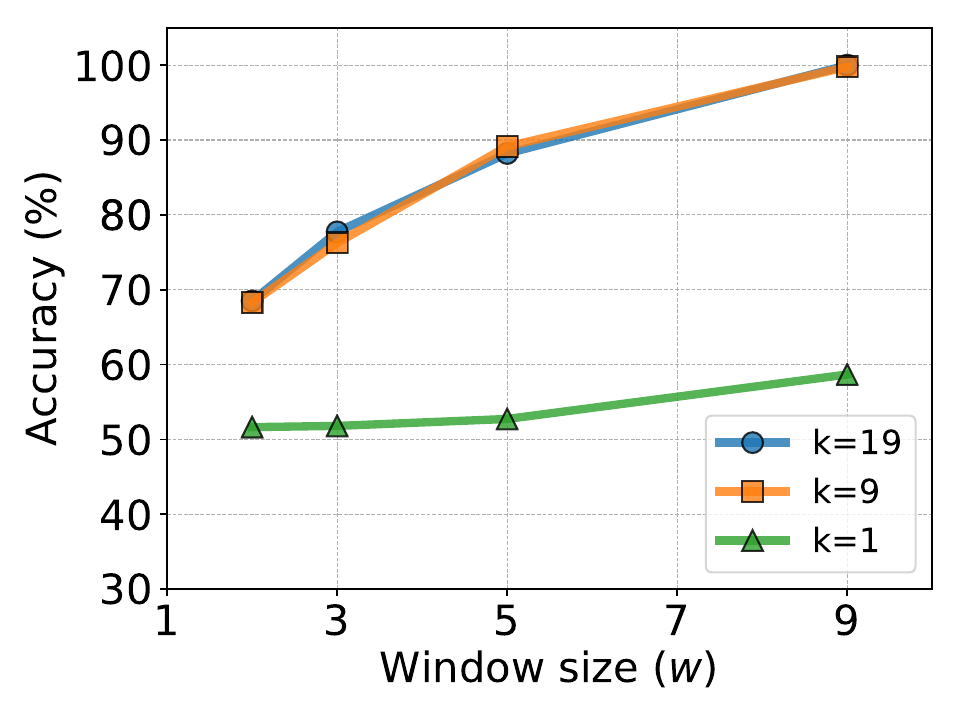}
        \caption{\f{mpi\_powm}. \label{fig:elgamal:alignment}}
    \end{subfigure}
    \hspace{-.2cm}
    \begin{subfigure}{0.24\textwidth}
        \centering
        \includegraphics[width=\textwidth]{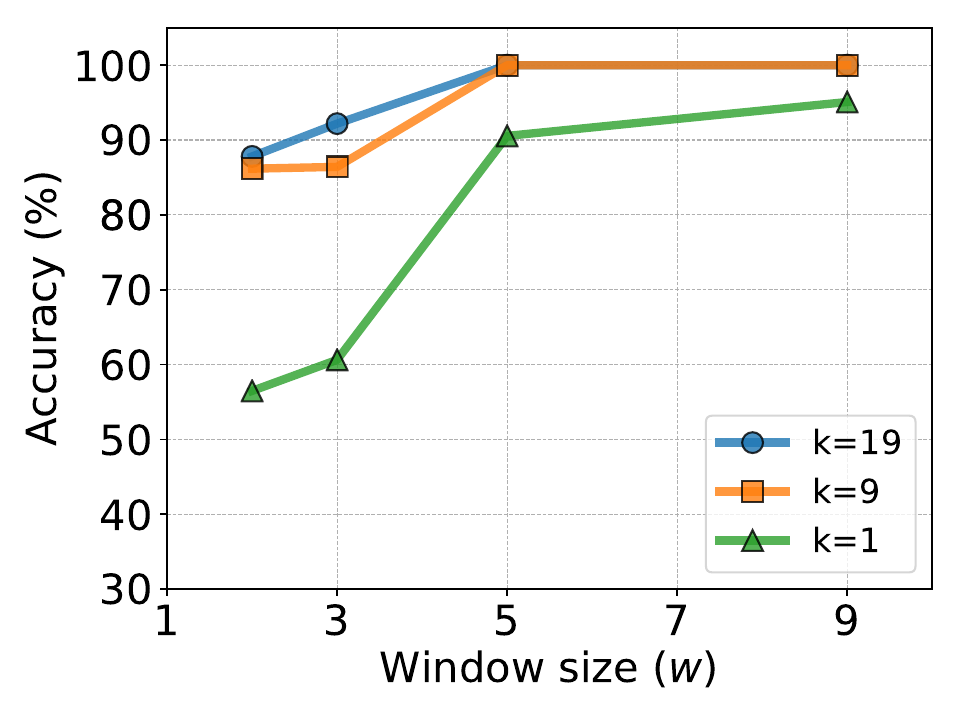}
        \caption{\f{mpi\_ec\_mul\_point}. \label{fig:eddsa:alignment}}
    \end{subfigure}%
\caption{Accuracy versus window-size for \CacheOnly.  Label ``k'' refers to the number of samples per segment. \label{fig:w}}
\vspace{-1 em}
\end{figure}

Column ``Attack Accuracy'' of Table~\ref{tab:elgamal:event} and Table~\ref{tab:eddsa:event} show that our attack strategy can recover between \pt{70} and \pt{100} of the enclave secret, depending on (i) the type of side-channel exploited, (ii) the detection tool used by the victim (either T-SGX or \ideal), and (iii) the number of consecutive code segments attacked per run for attacks that only exploit cache.
Our experiments also show that attack accuracy decreases when the victim is instrumented with T-SGX: this is likely due to the noise introduced by T-SGX when restarting transactions that abort before completion.

Comparing the accuracy of attacks on \f{mpi\_powm} with the accuracy when attacking \f{mpi\_ec\_mul\_point}, we note that \PageFault performs better on \f{mpi\_powm} and this is because the time difference between a \f{1}-loop and a \f{0}-loop in that routine is sharper than the time difference of the loops in \f{mpi\_ec\_mul\_point}. Nevertheless, attack variants that use cache are more accurate on \f{mpi\_ec\_mul\_point} as the cache side-channel is more noisy when attacking \f{mpi\_powm}. Furthermore, cache-only attack with larger windows (e.g., $w=9$) provide very good results.

We also assess the impact on accuracy of the number of samples $k$ we obtain for each secret bit. As shown in Figure~\ref{fig:k}, increasing $k$ improves accuracy that, however, plateaus around $k=9$ for most of the attack variants.

Finally, we assess the impact on accuracy when relying on a sliding window to reduce alignment errors in cache-only attacks. Recall that attacking a single segment at a time by only using cache side-channels may lead to poor results due to the difficulty of aligning the victim and attack threads (see Section~\ref{sec:attack}).
In our experiments, a cache-only attack on one segment at a time resulted in an average accuracy over $20$ runs of \avgstd{46.64\%}{3.84\%} for \f{mpi\_powm} with \ideal. The same experiment when attacking \f{mpi\_ec\_mul\_point}  showed an average accuracy of \avgstd{51.64\%}{1.98\%}. By using the sliding window technique described in Section~\ref{sec:attack}, we improve accuracy as shown in Figure~\ref{fig:w}. In particular, a window of size $w=9$ allows to recover the full secret when attacking \f{mpi\_powm}, whereas the same result can be achieved with a window of size $w=5$ for \f{mpi\_ec\_mul\_point}. This is because, the cache side-channel is less noisy in \f{mpi\_ec\_mul\_point}, as explained before.

\subsection{Effectiveness against detection tools} \label{eval:attack:defense}

\begin{figure*}[t]
    \centering
    \begin{subfigure}{.245\textwidth}
        \centering
        \includegraphics[width=\textwidth]{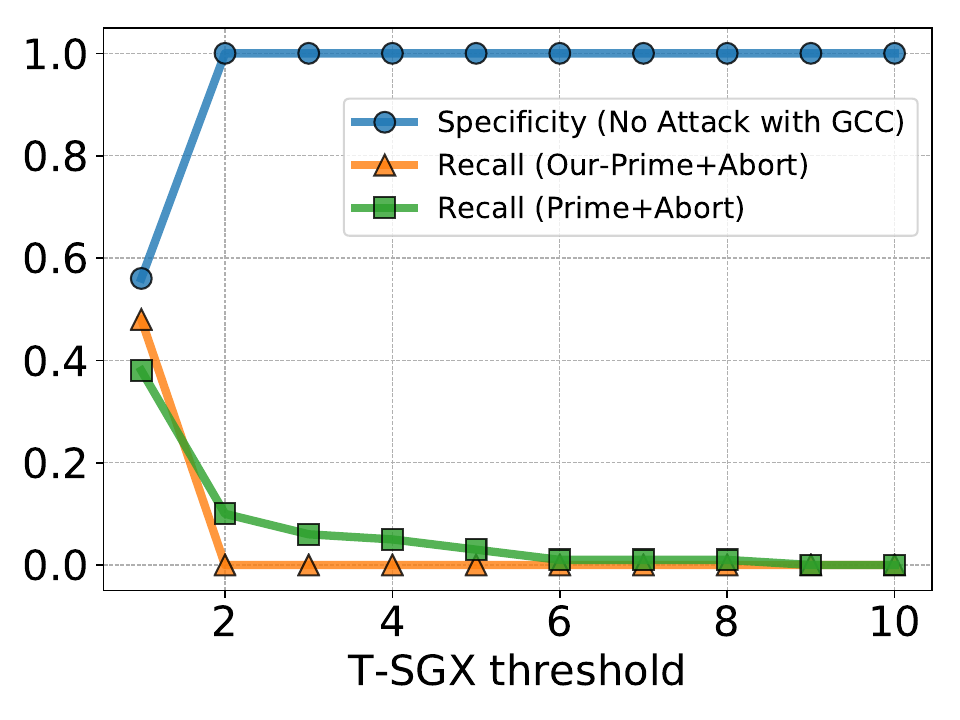}
        \caption{\f{mpi\_powm} \label{fig:T-SGX:elgamal:maxabort}}
    \end{subfigure}
    \begin{subfigure}{.245\textwidth}
        \centering
        \includegraphics[width=\textwidth]{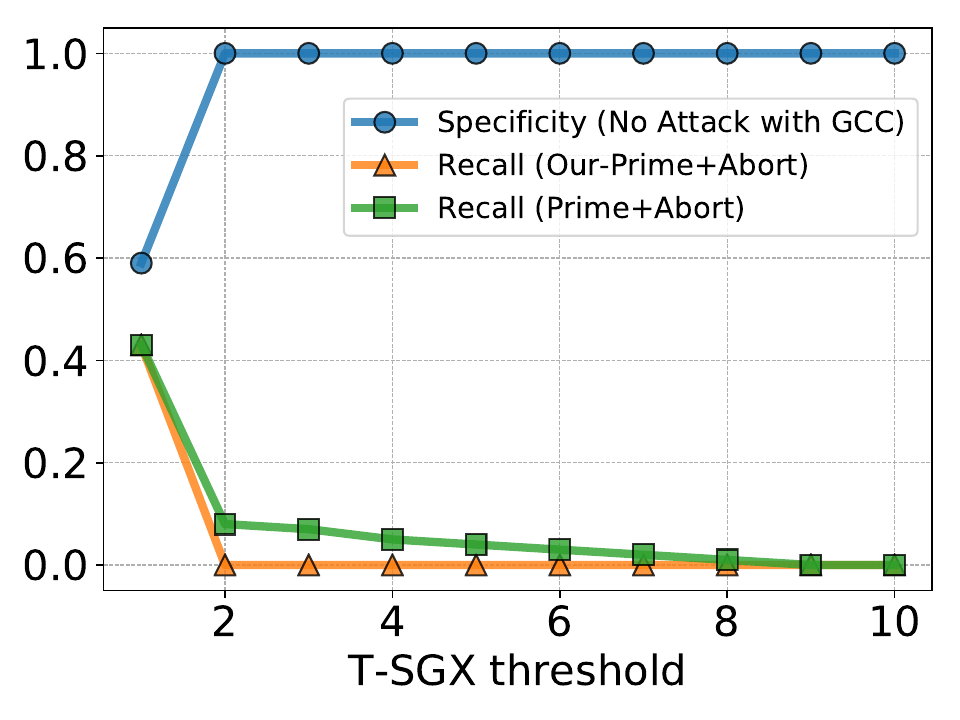}
        \caption{\f{mpi\_ec\_mul\_points} \label{fig:T-SGX:eddsa:maxabort}}
    \end{subfigure}
    \begin{subfigure}{.245\textwidth}
        \centering
        \includegraphics[width=\textwidth]{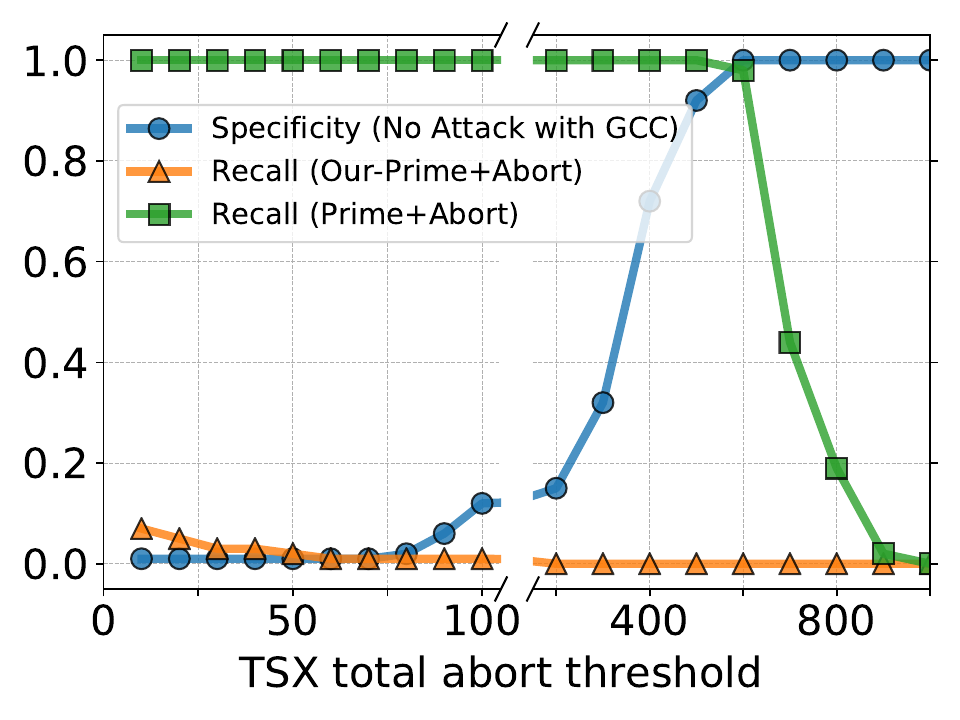}
        \caption{\f{mpi\_powm} \label{fig:T-SGX:elgamal}}
    \end{subfigure}
    \begin{subfigure}{.245\textwidth}
        \centering
        \includegraphics[width=\textwidth]{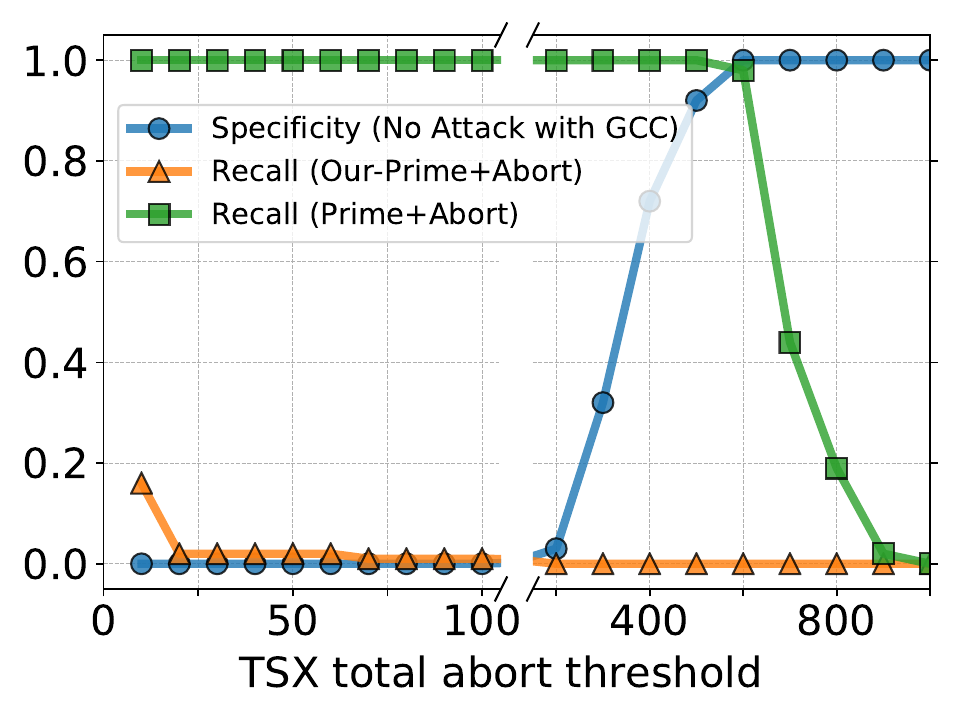}
        \caption{\f{mpi\_ec\_mul\_points} \label{fig:T-SGX:eddsa}}
    \end{subfigure}
    \caption{Specificity and recall of detection based on T-SGX when counting aborts per transaction ((a) and (b)) and when counting the total number of aborts across transactions ((c) and (d)). GCC is running on the same host. \label{fig:T-SGX} }
     \vspace{-1em}
\end{figure*}

\begin{figure*}[t]
    \centering
    \begin{subfigure}{.245\textwidth}
        \centering
        \includegraphics[width=\textwidth]{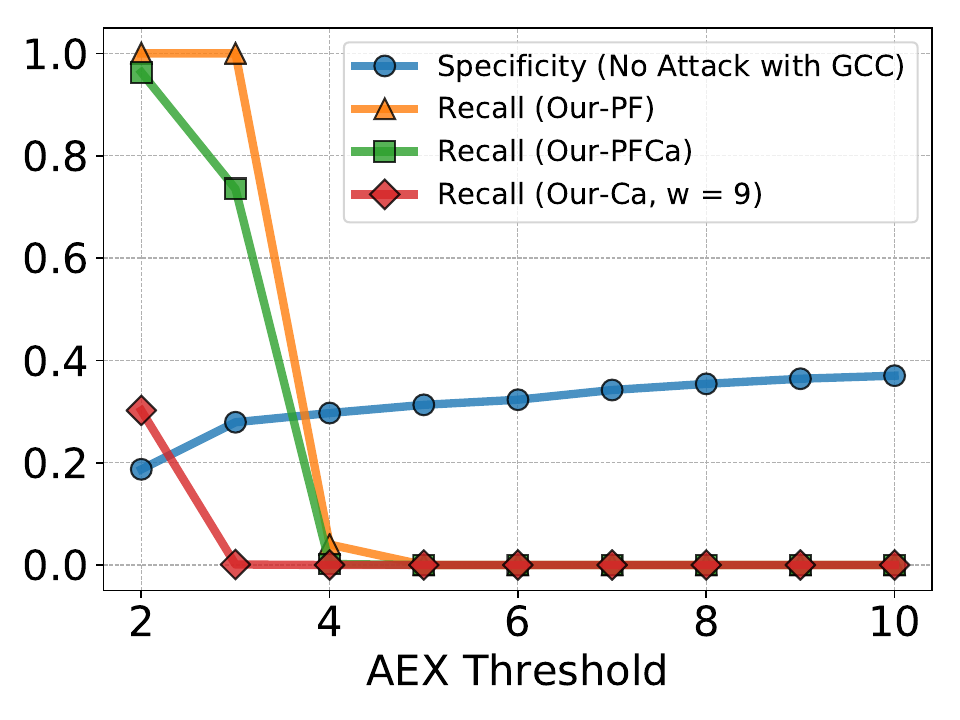}
        \caption{\f{mpi\_powm}  \label{fig:threshold:aex:elgamal}}
    \end{subfigure}
    \hspace{-.1cm}
    \begin{subfigure}{.245\textwidth}
        \centering
        \includegraphics[width=\textwidth]{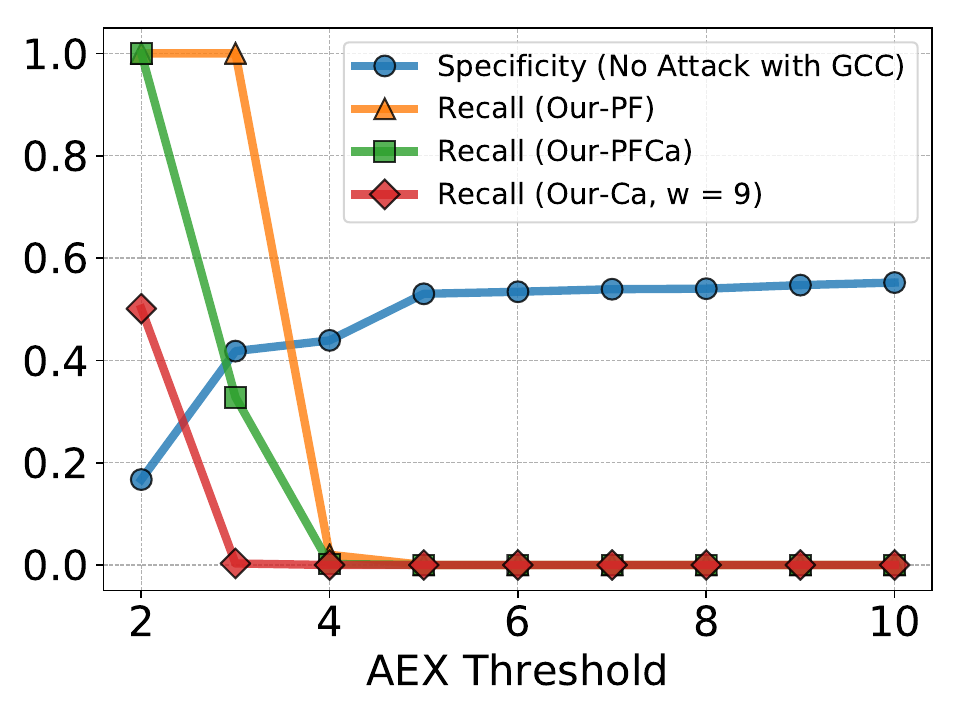}
        \caption{\f{mpi\_ec\_mul\_points}  \label{fig:threshold:aex:eddsa}}
    \end{subfigure}
    \begin{subfigure}{.245\textwidth}
        \centering
        \includegraphics[width=\textwidth]{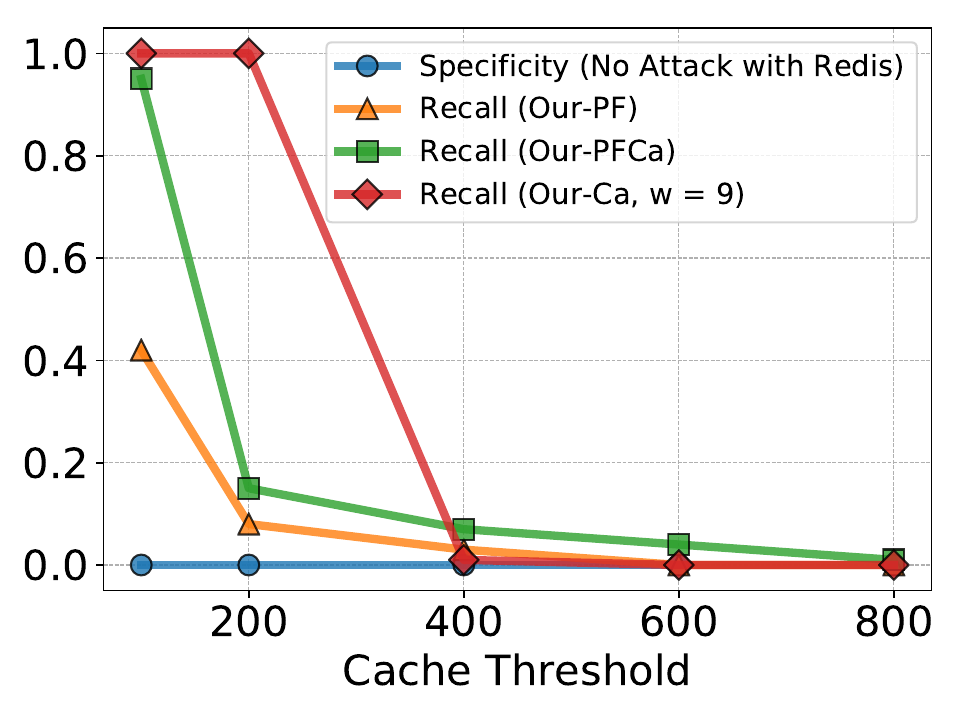}
        \caption{\f{mpi\_powm}  \label{fig:threshold:cache:elgamal}}
    \end{subfigure}
    \hspace{-.1cm}
    \begin{subfigure}{.245\textwidth}
        \centering
        \includegraphics[width=\textwidth]{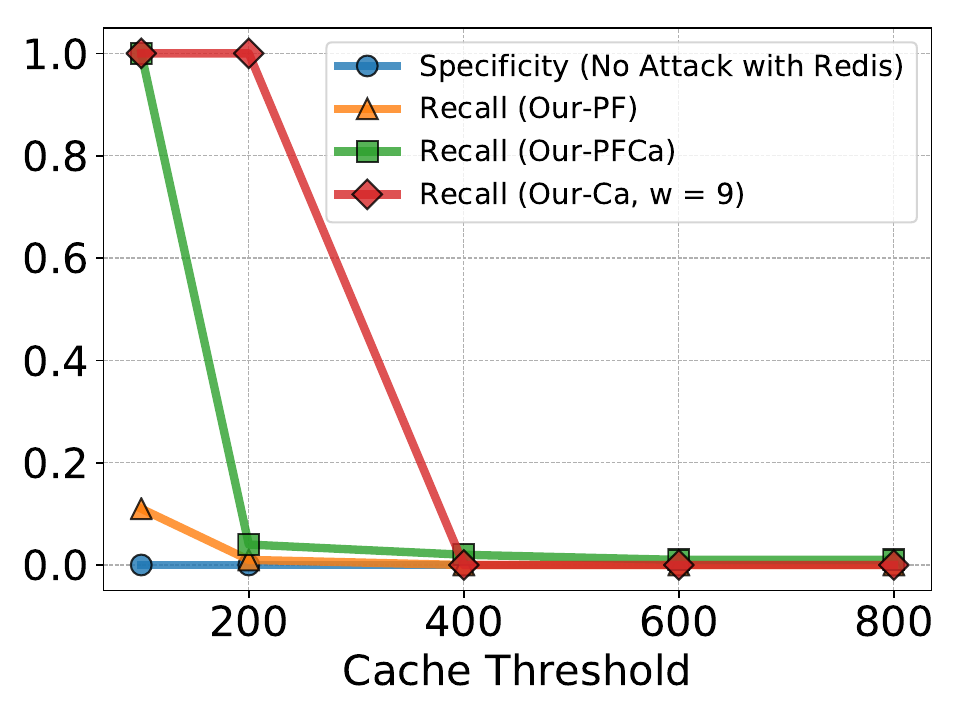}
        \caption{\f{mpi\_ec\_mul\_point}  \label{fig:threshold:cache:eddsa}}
    \end{subfigure}
\vspace{-0.5 em}
    \caption{Specificity and recall of detection based on \ideal.  \label{fig:ideal}}
    \vspace{-1em}
\end{figure*}

We now assess the effectiveness of T-SGX and \ideal{} described above in detecting an adversary using any of our attack strategies.  Recall that T-SGX monitors the number of transaction aborts whereas \ideal{} monitors performance metrics proposed in literature, namely number of AEXs, execution time, and cache misses.

We collect the aforementioned performance metrics, when the victim is under attack, as well as  when the victim is running in a benign environment either (i) alone or (ii) while another process is running on the same machine. For the latter, we used either \f{Redis}---a key-value store---and we mimic a realistic workload as in~\cite{redis-workload}, or GCC while building a large project. Finally, we record the required performance metrics while attacking the victim with standard side-channel attacks.

On the one hand, reported figures on routines instrumented with T-SGX provide us with evidence of the effectiveness of our attack strategies when the victim is equipped with existing tools. On the other hand, results of the experiments with \ideal{} allow us to reason about effectiveness of our attack strategies with respect to any tool that monitors the performance of the potential victim. For each scenario, we run the victim $1,000$ times and report the average and standard deviation of each metric in Table~\ref{tab:elgamal:event} and Table~\ref{tab:eddsa:event}. In particular, the tables show the number of AEX for \ideal{} and the total number of TSX aborts across transactions for T-SGX.

For each of the considered scenarios and for different values of the detection thresholds, we also  measure \emph{recall} (ratio of true positives over all positive cases) and \emph{specificity} (ratio of true negatives over all negative cases). The recall  metric is used to measure security: a perfect tool should have recall equal to one, and any smaller value means that the tool is not detecting some attacks. Specificity measures usability: a perfect tool should have specificity equal to one, and any smaller value means that the tool is raising some false alarms. In a real deployment, detection thresholds should be set so that both recall and specificity are as close as possible to one. In the following, we show that, for both T-SGX and \ideal,  high specificity and high recall cannot be achieved at the same time---which, in turn, shows that such tools cannot detect an adversary that uses our attack strategy.

\vspace{0.5 em}\noindent\textbf{Detection with T-SGX.}
In this set of experiments, we run the victim instrumented with T-SGX along with GCC to mimic a realistic multi-threaded workload. We vary the threshold number of transaction aborts before T-SGX raises an alarm, and measure specificity and recall for each threshold.

Figure~\ref{fig:T-SGX} (a) and (b) depicts the results for \f{mpi\_powm} and\\ \f{mpi\_ec\_mul\_point}. In order to reach a decent level of specificity (i.e., to avoid false-alarms), one should set the detection threshold $t\geq 2$. This result is inline with the experiments of the original T-SGX paper~\cite{tsgx:ndss17} that reports that most of the transactions of the applications abort and must be restarted up to two times before completing.
However, if $t\geq 2$, all of our attacks go undetected (i.e., recall is $0$). For completeness, Figure~\ref{fig:T-SGX} (a) and (b) also reports the recall value for the ``standard'' Prime+Abort attack of~\cite{prime:abort:sec17}: it is roughly $0.1$ for $t=2$ and decreases to $0$ when $t\geq 8$. In other words, T-SGX cannot detect Prime+Abort attacks. 
We note however, that Prime+Abort can be easily detected by T-SGX if it monitored the total number of transaction aborts, apart from the number of aborts per transaction. To show this, we have modified T-SGX to keep track of the number of aborts across all transactions and to raise an alarm if that reaches a specified threshold $t'$. Figure~\ref{fig:T-SGX} (c) and (d) shows specificity and recall for \f{mpi\_powm} and \f{mpi\_ec\_mul\_point}. It takes a threshold $t'$ of roughly 600 aborts to reach specificity close to one, in order to avoid false positives. Nevertheless, if $t'=600$, the standard TSX-based attack can be easily detected whereas \TSX goes unnoticed for $t'\geq 100$. Further, a Prime+Abort attack could be easily spotted by monitoring the victim's cache. In our experiments, Prime+Abort against \f{mpi\_powm} caused, on average a $\times 10$ increase of cache misses, compared to a benign execution. The increase of cache misses if the victim were running \f{mpi\_ec\_mul\_point} was $\times 19$ on average.

\vspace{0.5 em}\noindent\textbf{Detection with \ideal.}
In case of \ideal, we consider detection based on both the number of AEXs and the number of cache misses. For each scenario, we run the victim either along with GCC---to mimic a multi-threaded workload---or along with Redis---as an exemplary application to mimic a memory-intensive workload.
Further, we vary the detection threshold---either the one of number of AEXs or the one of number of cache misses--- and measure specificity and recall for each threshold.

Figure~\ref{fig:threshold:aex:elgamal} and Figure~\ref{fig:threshold:aex:eddsa} show results when the tool is monitoring AEXs and the victims are \f{mpi\_powm} or \f{mpi\_ec\_mul\_point}, respectively. No threshold in the range we consider ($2\leq t\leq 10$) provides specificity greater than $0.37$ in case of \f{mpi\_powm} and $0.55$ in case of \f{mpi\_ec\_mul\_point}. At the same time, all of the attack variants go undetected (recall is $0$) if $t\geq 5$ for both victims.

Figure~\ref{fig:threshold:cache:elgamal} and Figure~\ref{fig:threshold:cache:eddsa} depict our results when the tool is monitoring cache misses and the victims are \f{mpi\_powm} or\\ \f{mpi\_ec\_mul\_point}, respectively. For both victims, specificity is $0$ for thresholds up to $800$; hence, one should set a much higher threshold to avoid false alarms. However, all of the attacks go undetected if $t\geq 800$ for \f{mpi\_powm} or $t\geq 600$ for \f{mpi\_ec\_mul\_point}.

\vspace{0.5 em}\noindent\textbf{Detection by monitoring execution time.} A detection tool may monitor the execution time to decide whether the application is under attack. This is for example the case of \dejavu~\cite{chen17asiaccs}. Results from Table~\ref{tab:elgamal:event} and Table~\ref{tab:eddsa:event} show that standard page-fault attacks almost double the execution time and would be likely detected by tools such as \dejavu. Differently, our attacks cause minimal increase of the victim's execution time (below \pt{2}); as such, it is challenging for \dejavu or similar tools to detect them.

\subsection{T-SGX with cache-based monitoring} \label{appendix:tsgx:ideal}

So far, we have not considered cache-based attacks against T-SGX. This is because T-SGX~\cite{tsgx:ndss17} does not monitor the cache performance of an enclave and considers cache-based attacks as out of scope. In this section, we show that even if T-SGX were to be enhanced with additional functionality from \ideal{} (such as monitoring cache performance), it would still not be able to detect our fine-grained attacks.

If T-SGX were to be enhanced with the performance monitoring of \ideal, the result is a detection mechanism that (i) uses TSX to suppress page-faults notification to the OS, (ii) keeps track of the number of aborts per transaction, and (iii) monitors the number of cache misses. We then use cache-based attacks against this enhanced version of T-SGX and assess whether it can distinguish attacks from benign runs.

Table~\ref{tab:elgamal:event:appendix} and Table~\ref{tab:eddsa:event:appendix} summarize our findings for \f{mpi\_powm} and \f{mpi\_ec\_mul\_point}. As expected, \TSXCache performs slightly worse than \PageCache (e.g., $~70\%$ vs $~60\%$ accuracy for \f{mpi\_powm}): this is because accuracy of stopping at a specific code segment an enclave is higher if the enclave exposes page-faults. We also note that attack strategies that only leverage cache can recover the whole secrets with alignment windows of size $w=9$.

\begin{figure}[t]
    \centering
    \begin{subfigure}{0.3\textwidth}
        \centering
        \includegraphics[width=\textwidth]{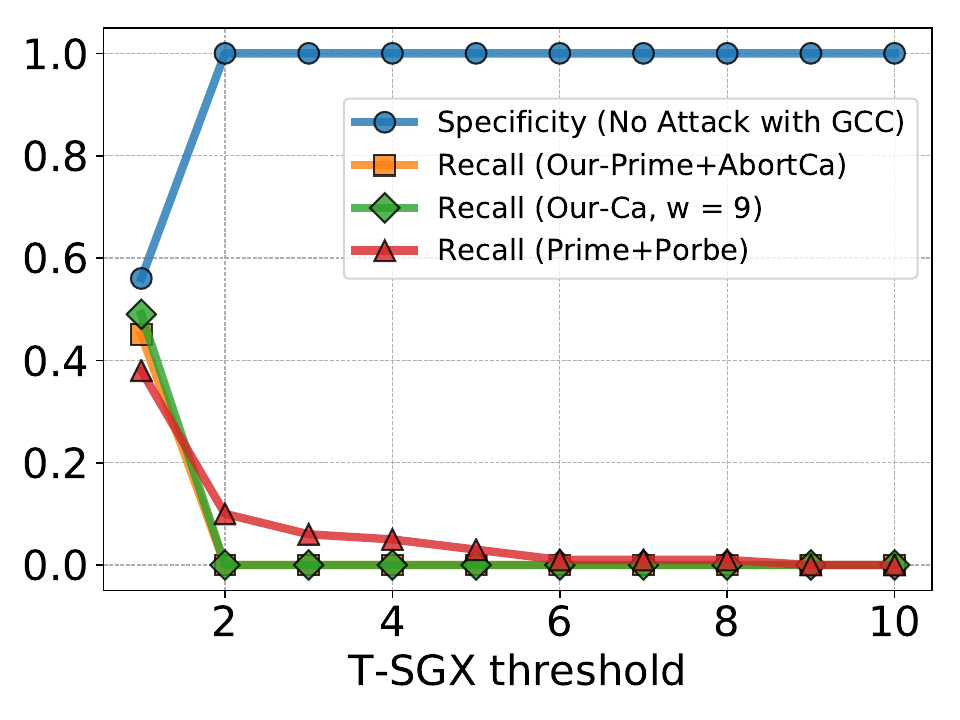}
        \vspace{-1em}
        \caption{\f{mpi\_powm} \label{fig:T-SGX:elgamal:maxabort:cache}}
        \vspace{1em}
    \end{subfigure}
    \begin{subfigure}{0.3\textwidth}
        \centering
        \includegraphics[width=\textwidth]{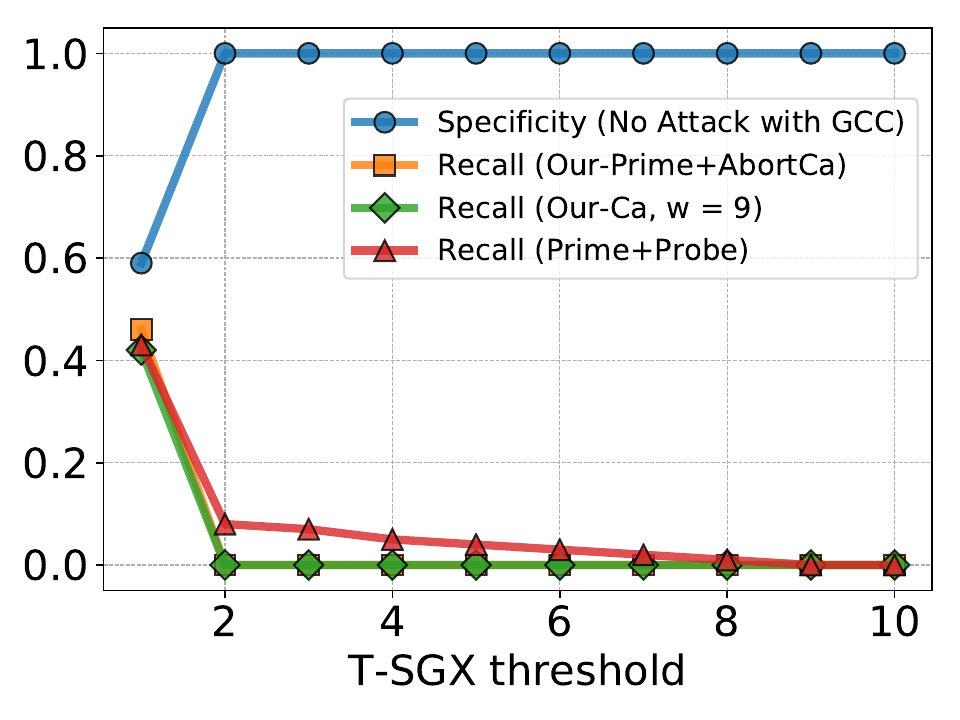}
        \caption{\f{mpi\_ec\_mul\_points} \label{fig:T-SGX:eddsa:maxabort:cache}}
    \end{subfigure}
    \caption{Specificity and recall of detection based on T-SGX when GCC is running on the same host. \label{fig:tsgx:threshold:cache} }
    \vspace{-1em}
\end{figure}
\begin{figure}[tbp]
    \centering
    \begin{subfigure}{0.3\textwidth}
        \centering
        \includegraphics[width=\textwidth]{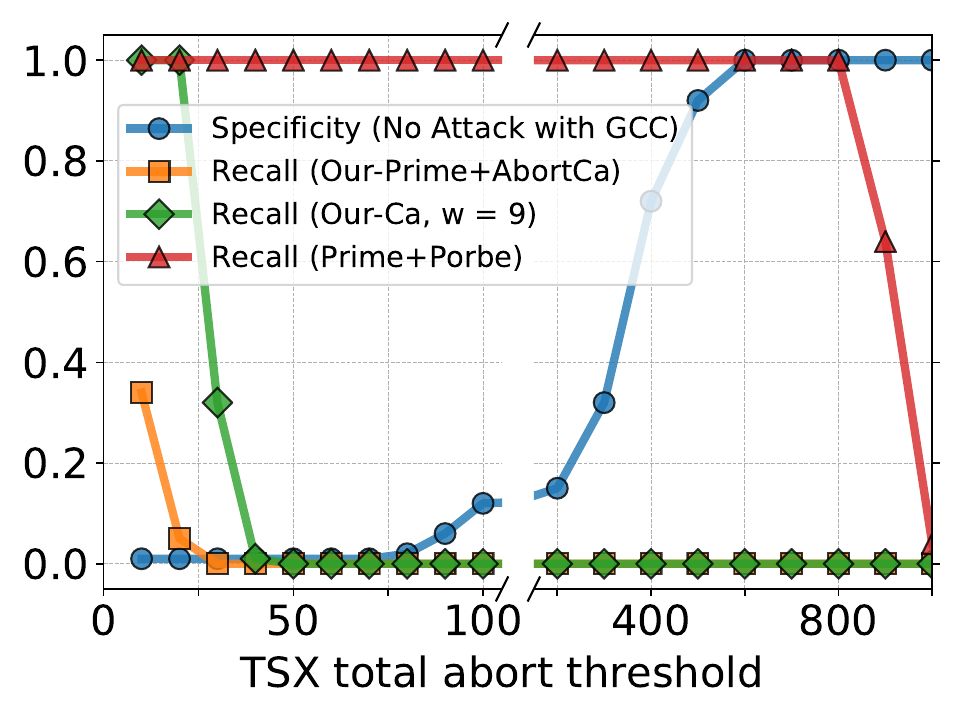}
        \vspace{-1em}
        \caption{\f{mpi\_powm} \label{fig:T-SGX:elgamal}}
        \vspace{1em}
    \end{subfigure}
    \begin{subfigure}{0.3\textwidth}
        \centering
        \includegraphics[width=\textwidth]{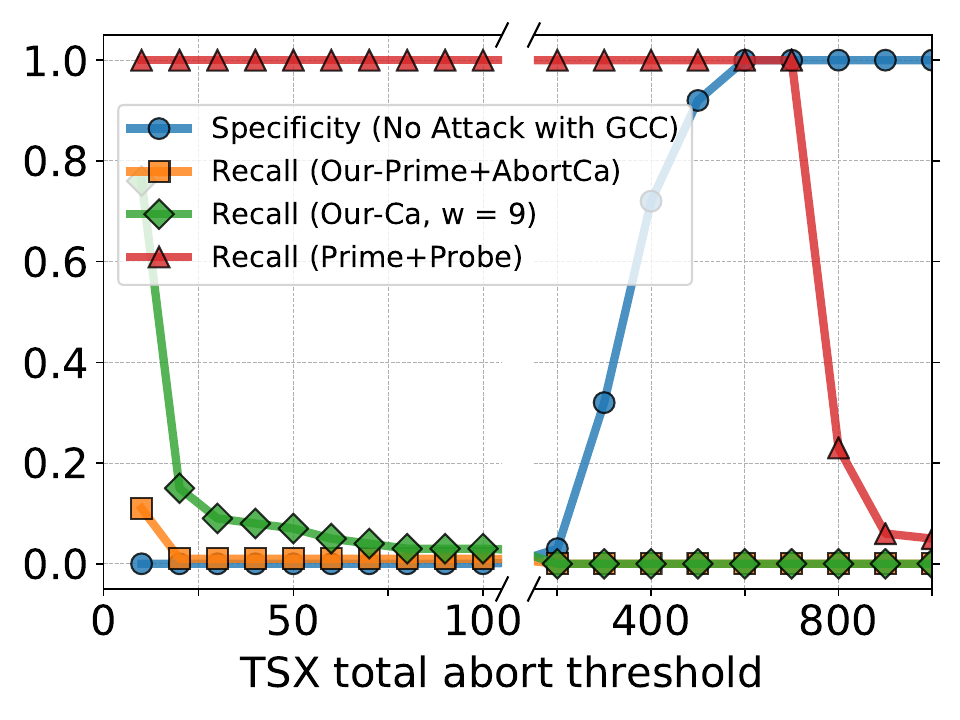}
        \caption{\f{mpi\_ec\_mul\_points} \label{fig:T-SGX:eddsa}}
    \end{subfigure}
    \caption{Specificity and recall of detection based when the total number of aborts across transactions is used in T-SGX. GCC is running on the same host.  \label{fig:tsgx:maxabort:cache}}
\vspace{-1.5 em}
\end{figure}

Regarding performance metric, we note that a standard cache attack noticeably increases the number of cache misses at the victim compared to benign run when, for example, GCC is running on the same host (between $\times 10$ and $\times 20$) whereas \CacheOnly causes a number of cache misses at the victim that is comparable to a benign run with no other application on the same platform.

Finally, Figure~\ref{fig:tsgx:threshold:cache} and Figure~\ref{fig:tsgx:maxabort:cache} show precision and recall of T-SGX and the enhanced T-SGX detection tool against cache-based attacks, respectively. Figure~\ref{fig:tsgx:threshold:cache} shows that the threshold of allowed aborts per transaction should be at least 2, in order to have specificity close to 1 (so to avoid false positives). In this case, all cache-based approaches are likely to go undetected (recall close to 0); we note that a standard ``Prime+Probe'' attack (recall around 0.1) performs slightly worse than our attacks (recall close to 0).

If T-SGX were modified to also monitor the total number of transaction aborts, Figure~\ref{fig:tsgx:maxabort:cache} suggests that a threshold of at least $600$ must be used to achieve precision close to 1 and avoid false positives. With this threshold, a standard ``Prime+Probe'' attack can be detected (recall is 1) while our cache-based attacks remain unnoticed (recall is 0).

\begin{table*}[!htb]
    \center
     \scalebox{1.0}{
\begin{tabular}{c|c|c|c|c|c}

&  &\textbf{Attack Accuracy}& \textbf{TSX Aborts}      & \textbf{L3 Cache-misses} & \textbf{Time (ms)} \\

\hline
\hline

\multirow{4}{*}{Our attacks} &\TSXCache& \avgstd{\pt{60.2}}{~\pt{3.4}} & \avgstd{9.56}{3.85}   & \avgstd{247.69}{153.40} & \avgstd{5.71}{0.02} \\

    &\CacheOnly ($w = 3$) &\avgstd{\pt{76.10}}{~\pt{10.52}}& \avgstd{7.11}{18.46}   & \avgstd{237.06}{115.11}     & \avgstd{5.66}{0.06}  \\

    &\CacheOnly ($w = 5$) &\avgstd{\pt{86.42}}{~\pt{14.32}} & \avgstd{18.59}{1.01}    & \avgstd{284.94}{69.83}    & \avgstd{5.68}{0.05} \\

    &\CacheOnly ($w = 9$) &\pt{100}& \avgstd{28.16}{3.74}   & \avgstd{388.20}{87.73}   & \avgstd{5.67}{0.04} \\

\hline
\hline

\multirow{1}{*}{Standard attacks}&\CacheAttack &\avgstd{\pt{84.4}}{~\pt{9.6}}& \avgstd{923.18}{55.52}       & \avgstd{3697.68}{817.95}   &  \avgstd{5.81}{0.013} \\

\hline
\hline

\multirow{3}{*}{No attack}    &\f{mpi\_powm}  && \avgstd{6.10}{21.76}                          & \avgstd{416.48}{410.76}                           & \avgstd{5.66}{0.02}  \\

    &\f{mpi\_powm} (w/ GCC)&  & \avgstd{188.67}{85.14} & \avgstd{394.14}{549.7}    & \avgstd{5.71}{0.60} \\

    &\f{mpi\_powm} (w/ Redis)& & \avgstd{121.25}{120.68}                  & \avgstd{16256.47}{5843.78}              & \avgstd{6.12}{0.25} \\

\hline
\hline

    \end{tabular}
     }
    \caption{Accuracy and performance metrics for \f{mpi\_powm}.}
    \label{tab:elgamal:event:appendix}
    \vspace{-1.5 em}
\end{table*}

\begin{table*}[!htb]
    \center
    \scalebox{0.925}{
        \begin{tabular}{c|c|c|c|c|c}
    & &\textbf{Attack Accuracy} &  \textbf{TSX Aborts}      & \textbf{L3 Cache-misses} & \textbf{Time (ms)} \\
\hline
\hline

\multirow{4}{*}{Our attacks} &\TSXCache &\avgstd{\pt{55.3}}{\pt{2.5}}& \avgstd{6.06}{15.54}   & \avgstd{303.66}{344.21}       & \avgstd{13.95}{0.40}   \\

&\CacheOnly ($w = 3$)       & \avgstd{\pt{61.44}}{~\pt{14.1}} & \avgstd{17.65}{42.48}   & \avgstd{268.35}{562.07}       & \avgstd{13.45}{0.26}      \\

&\CacheOnly ($w = 5$)    & \avgstd{\pt{75.91}}{~\pt{10.1}} & \avgstd{18.73}{43.10}   & \avgstd{257.92}{448.48}       & \avgstd{13.70}{0.32}        \\

&\CacheOnly ($w = 9$)  & \pt{100} & \avgstd{23.9}{55.37}   & \avgstd{305.34}{524.03}        & \avgstd{13.36}{0.24}        \\

\hline
\hline

\multirow{1}{*}{Standard attacks} &\CacheAttack                      &\avgstd{\pt{96.6}}{\pt{3.8}}& \avgstd{788.89}{67.50}   &  \avgstd{15730.48}{4363.18}       &  \avgstd{14.17}{0.19}        \\

\hline
\hline

\multirow{3}{*}{No attack}&\f{mpi\_ec\_mul\_points}    &   & \avgstd{6.22}{34.48}   & \avgstd{158.93}{133.85}       & \avgstd{13.31}{0.29}          \\

&\f{mpi\_ec\_mul\_points} (w/ GCC)  &   & \avgstd{377.93}{87.54} & \avgstd{817.54}{1331.89}    & \avgstd{13.67}{0.50} \\

&\f{mpi\_ec\_mul\_points} (w/ Redis)      &   & \avgstd{226.91}{109.03}   & \avgstd{84941.38}{25664.85}       & \avgstd{17.69}{1.17}               \\

\hline
\hline

    \end{tabular}
    }
    \caption{Accuracy and performance metrics for \f{mpi\_ec\_mul\_point}.
    }
    \label{tab:eddsa:event:appendix}
    \vspace{-2 em}
\end{table*}

\begin{table*}[t]
    \footnotesize
    \center
    \scalebox{1.1}{\begin{tabular}{l|c|c|c|c|c}
                            &  & \textbf{Attack Accuracy} & \textbf{AEX} & \textbf{L3 Cache-misses} & \textbf{Time (ms)} \\
        \hline
        \hline
        \multirow{2}{*}{No attack} & \f{DTreesImpl::predictTrees}  & & \avgstd{7.83}{0.49}   & \avgstd{134.55}{85.38} & \avgstd{2.46}{0.06} \\
                                    & \f{DTreesImpl::predictTrees} (w/ GCC)                     &                       & \avgstd{16.74}{1.44}   & \avgstd{132.42}{73.18} & \avgstd{2.55}{0.03} \\
        \hline
        Standard attack            & Page-faults attack    & \avgstd{\pt{65.2}}{0} & \avgstd{3070.9}{1.4}    & \avgstd{2164}{5455.22} & \avgstd{58.81}{0.41} \\
        \hline
        Our attack            & \PageFault            & \avgstd{\pt{54.9}}{~\pt{3.61}}    & \avgstd{8.21}{1.34}     & \avgstd{150.77}{74.67} & \avgstd{2.47}{0.08} \\
        \hline
        \hline
    \end{tabular}}
    \caption{Accuracy and performance metrics for \f{DTreesImpl::predictTrees} with \ideal.}
    \vspace{-1.5 em}
    \label{tab:opencv:event}
\end{table*}

\section{Attacking OpenCV}
\label{sec:opencv}

We now adapt our attack strategy to known side-channels of decision-tree routines~\cite{oblivious:sec16} to assess the feasibility of attacking the decision-tree routine of OpenCV~\cite{opencv}---a well-known computer vision library. Similar to previous work~\cite{kaggle:tree:mnist}, we use the MNIST~\cite{mnist} data-set and assume an application consisting of an enclaved execution of OpenCV's decision-trees to detect handwritten digits.

The decision-tree traversal function of OpenCV walks the tree and, depending on the input image, accesses different nodes, resulting in different page accesses. We use page-faults to infer the pattern of page accesses and leak the prediction output. To capture the access pattern of different input images, we rely on an offline analysis of the routine and observe memory page access patterns. Thus, we set such memory pages as fault during runtime, to infer the prediction output.

For training the decision-tree, we rely on $60,000$ samples from the MNIST data set. During the inference phase, the trained decision-tree model is first loaded into the enclave and then used to recognize $100$ input images at a time from a set of $10,000$ test images.
The execution time of image recognition is almost independent of the input image, as the tree is almost balanced. Therefore, we can model the execution time as $T_{i} = T_i + c$, where $c$ is a constant value. In our experiments, we found out that $c$ is roughly $9.7$k clock ticks ($\sigma = 975.3$).

\noindent\textbf{Profiling  and attack accuracy.} We successfully stopped the enclave at the $i$-th invocation of \f{DTreesImpl::predictTrees}  around 8 out of 10 times (\avgstd{~\pt{84.8}}{~\pt{7.85}}). The corresponding accuracy is reported in Table~\ref{tab:opencv:event}. We observe that a variant attack leveraging page-faults (\PageFault) is only slightly less accurate than a standard page-fault attack.

\begin{figure}[tb]
    \centering
    \includegraphics[width=.36\textwidth]{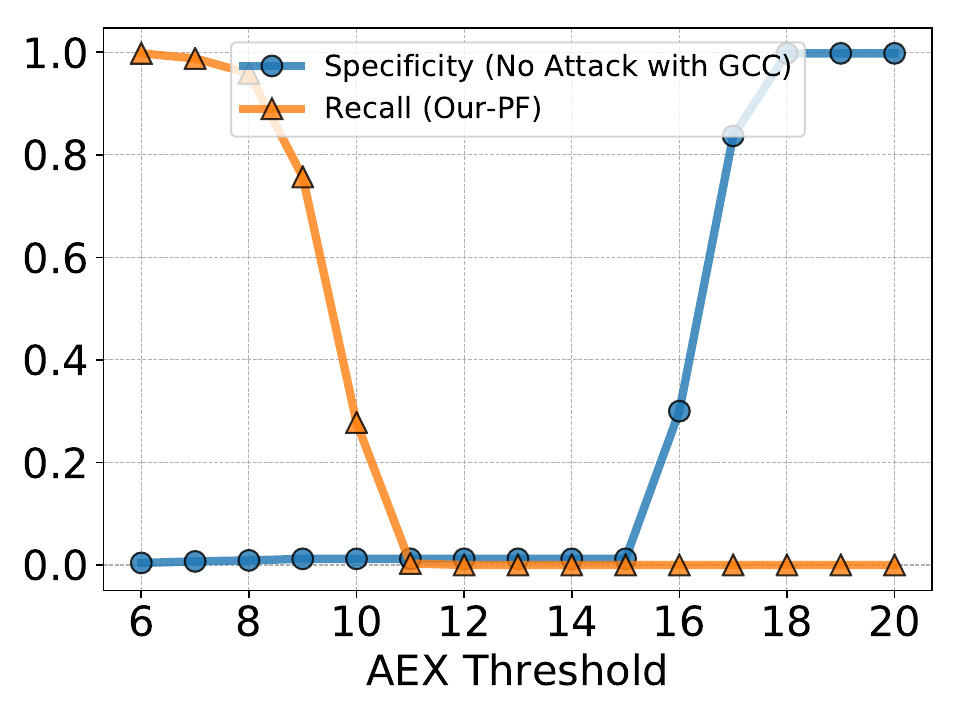}
    \vspace{-1 em}
    \caption{Specificity and recall of detection based on the number of AEXs when GCC is running on the same host.}
    \vspace{-1em}
    \label{fig:threshold:opencv}
\end{figure}

\noindent\textbf{Effectiveness against detection tools.} To analyze the effectiveness of our strategy against detection tools, we measure specificity and recall for different AEX thresholds. We only assume the victim is equipped with \ideal---as T-SGX supports only C, we could not instrument OpenCV using T-SGX.
Figure~\ref{fig:threshold:opencv} shows that no threshold value can achieve high specificity and high recall at the same time. In particular, if the detection threshold is smaller than $17$, then specificity falls below $0.84$ (i.e., a false alarm is raised 2 out of 10 times). At the same time, a detection threshold equal to or bigger than $11$ allows attacks to go undetected (recall=$0.003$). We also note that \PageFault causes no noticeable overhead in terms of cache misses or execution time. We conclude that \ideal---monitoring number of AEXs, cache misses or execution time---may not be able to tell an attack that uses our strategy from a benign run of the victim enclave.

\section{Concluding remarks} \label{sec:discuss}

In this paper, we analyzed the limitations of existing detection tools that monitor performance metrics to detect side-channel attacks on SGX enclaves. Our findings show that an adversary can bypass existing tools when deployed in realistic cloud scenarios by exfiltrating small portions of a secret at each run of the victim. 

One possible countermeasure would be to instruct detection tools to keep \emph{state} to detect a pattern of small anomalies spread across multiple executions. Intel SGX, however, does not provide freshness of state information sealed to disk. A malicious OS can, therefore, easily bypass such a tool by providing stale state to the enclave. Another countermeasure could be to prevent arbitrary restarts of the victim enclave by, e.g., programming the enclave to run only upon receiving an authenticated request.
Nevertheless, this option is hardly workable when the enclave provides a ``public'' service. For instance, if the enclave hosts a TLS server~\cite{LibSeal} or a password-hardening service~\cite{safekeeper}, it is extremely challenging to differentiate between an authorized request from a honest user and another issued by the adversary acting as a honest user.

We hope that our findings will motivate further research in this area, with the aim to avoid unnecessary cycles of attacks/defenses on detection tools that solely rely on performance metrics.

\begin{acks}
  We thank all anonymous reviewers for their helpful comments. This work was partially funded by the European Union’s Horizon 2020 research and innovation programme under grant agreement No. 957406 (TERMINET), and by the Deutsche Forschungsgemeinschaft (DFG, German Research Foundation) under Germany’s Excellence Strategy - EXC 2092 CASA - 390781972.
\end{acks}

\bibliographystyle{plain}
\bibliography{biblio}

\begin{thebibliography}{10}

\bibitem{kaggle:tree:mnist}
97\% on {MNIST} with a single decision tree (+ t-sne).
\newblock
  \url{https://www.kaggle.com/carlolepelaars/97-on-mnist-with-a-single-decision-tree-t-sne}.

\bibitem{mnist}
Mnist dataset.
\newblock \url{http://yann.lecun.com/exdb/mnist/}.

\bibitem{opencv}
Opencv.
\newblock \url{https://github.com/opencv/opencv}.

\bibitem{redis-workload}
Redis benchmark.
\newblock \url{https://redis.io/topics/benchmarks}.

\bibitem{obfuscuro:ndss19}
Adil Ahmad, Byunggill Joe, Yuan Xiao, Yinqian Zhang, Insik Shin, and
  Byoungyoung Lee.
\newblock Obfuscuro: A commodity obfuscation engine on intel sgx.
\newblock In {\em NDSS}, 2019.

\bibitem{LibSeal}
Pierre{-}Louis Aublin, Florian Kelbert, Dan O'Keeffe, Divya Muthukumaran,
  Christian Priebe, Joshua Lind, Robert Krahn, Christof Fetzer, David~M. Eyers,
  and Peter~R. Pietzuch.
\newblock Libseal: revealing service integrity violations using trusted
  execution.
\newblock In {\em Proceedings of the Thirteenth EuroSys Conference, {EuroSys}},
  pages 24:1--24:15, 2018.

\bibitem{sgx:clock}
Maurice Bailleu, Donald Dragoti, Pramod Bhatotia, and Christof Fetzer.
\newblock Tee-perf: A profiler for trusted execution environments.
\newblock In {\em 2019 49th Annual IEEE/IFIP International Conference on
  Dependable Systems and Networks (DSN)}, pages 414--421. IEEE, 2019.

\bibitem{drsgx:acsac19}
Ferdinand Brasser, Srdjan Capkun, Alexandra Dmitrienko, Tommaso Frassetto, Kari
  Kostiainen, and Ahmad-Reza Sadeghi.
\newblock Dr.sgx: Automated and adjustable side-channel protection for sgx
  using data location randomization.
\newblock In {\em Proceedings of the 35th Annual Computer Security Applications
  Conference}, ACSAC '19, pages 788--800, New York, NY, USA, 2019. ACM.

\bibitem{brasser17woot}
Ferdinand Brasser, Urs M{\"{u}}ller, Alexandra Dmitrienko, Kari Kostiainen,
  Srdjan Capkun, and Ahmad{-}Reza Sadeghi.
\newblock Software grand exposure: {SGX} cache attacks are practical.
\newblock In {\em {USENIX} Workshop on Offensive Technologies ({WOOT})}, pages
  1--12, 2017.

\bibitem{briongos18codaspy}
Samira Briongos, Gorka Irazoqui, Pedro Malag{\'{o}}n, and Thomas Eisenbarth.
\newblock Cacheshield: Detecting cache attacks through self-observation.
\newblock In Ziming Zhao, Gail{-}Joon Ahn, Ram Krishnan, and Gabriel Ghinita,
  editors, {\em Proceedings of the Eighth {ACM} Conference on Data and
  Application Security and Privacy, {CODASPY} 2018, Tempe, AZ, USA, March
  19-21, 2018}, pages 224--235. {ACM}, 2018.

\bibitem{stealth:page:security17}
Jo~Van Bulck, Nico Weichbrodt, R{\"u}diger Kapitza, Frank Piessens, and Raoul
  Strackx.
\newblock Telling your secrets without page faults: Stealthy page table-based
  attacks on enclaved execution.
\newblock In {\em 26th {USENIX} Security Symposium ({USENIX} Security 17)},
  pages 1041--1056, 2017.

\bibitem{hyperrace:sp18}
G.~{Chen}, W.~{Wang}, T.~{Chen}, S.~{Chen}, Y.~{Zhang}, X.~{Wang}, T.~{Lai},
  and D.~{Lin}.
\newblock Racing in hyperspace: Closing hyper-threading side channels on sgx
  with contrived data races.
\newblock In {\em 2018 IEEE Symposium on Security and Privacy (SP)}, pages
  178--194, May 2018.

\bibitem{dejavurepo}
Sanchuan Chen.
\newblock {D{\'{e}}j{\`{a}} Vu}.
\newblock \url{https://github.com/schuan/dejavu}.
\newblock Accessed on 22/11/2021.

\bibitem{chen17asiaccs}
Sanchuan Chen, Xiaokuan Zhang, Michael~K. Reiter, and Yinqian Zhang.
\newblock Detecting privileged side-channel attacks in shielded execution with
  d{\'{e}}j{\`{a}} vu.
\newblock In {\em {ACM} Asia Conference on Computer and Communications
  Security, ({AsiaCCS})}, pages 7--18, 2017.

\bibitem{prime:abort:sec17}
Craig Disselkoen, David Kohlbrenner, Leo Porter, and Dean~M. Tullsen.
\newblock Prime+abort: {A} timer-free high-precision {L3} cache attack using
  intel {TSX}.
\newblock In {\em 26th {USENIX} Security Symposium, {USENIX Security}}, pages
  51--67, 2017.

\bibitem{cloak:security17}
Daniel Gruss, Julian Lettner, Felix Schuster, Olya Ohrimenko, Istvan Haller,
  and Manuel Costa.
\newblock Strong and efficient cache side-channel protection using hardware
  transactional memory.
\newblock In {\em 26th {USENIX} Security Symposium ({USENIX} Security 17)},
  pages 217--233, Vancouver, BC, August 2017. {USENIX} Association.

\bibitem{safekeeper}
Arseny Kurnikov, Klaudia Krawiecka, Andrew Paverd, Mohammad Mannan, and
  N.~Asokan.
\newblock Using safekeeper to protect web passwords.
\newblock In {\em The Web Conference, {WWW}}, pages 159--162, 2018.

\bibitem{llcattack:sp15}
Fangfei Liu, Yuval Yarom, Qian Ge, Gernot Heiser, and Ruby~B Lee.
\newblock Last-level cache side-channel attacks are practical.
\newblock In {\em 2015 IEEE symposium on security and privacy}, pages 605--622.
  IEEE, 2015.

\bibitem{reverse:llc:raid15}
Cl{\'e}mentine Maurice, Nicolas Le~Scouarnec, Christoph Neumann, Olivier Heen,
  and Aur{\'e}lien Francillon.
\newblock Reverse engineering intel last-level cache complex addressing using
  performance counters.
\newblock In {\em International Symposium on Recent Advances in Intrusion
  Detection}, pages 48--65. Springer, 2015.

\bibitem{moghimi17ches}
Ahmad Moghimi, Gorka Irazoqui, and Thomas Eisenbarth.
\newblock Cachezoom: How {SGX} amplifies the power of cache attacks.
\newblock In {\em International Conference on Cryptographic Hardware and
  Embedded Systems ({CHES})}, pages 69--90, 2017.

\bibitem{copycat:sec20}
Daniel Moghimi, Jo~Van~Bulck, Nadia Heninger, Frank Piessens, and Berk Sunar.
\newblock {CopyCat}: Controlled instruction-level attacks on enclaves.
\newblock In {\em 29th {USENIX} Security Symposium}, pages 469--486, August
  2020.

\bibitem{oblivious:sec16}
Olga Ohrimenko, Felix Schuster, C{\'e}dric Fournet, Aastha Mehta, Sebastian
  Nowozin, Kapil Vaswani, and Manuel Costa.
\newblock Oblivious multi-party machine learning on trusted processors.
\newblock In {\em 25th {USENIX} Security Symposium ({USENIX} Security 16)},
  pages 619--636, 2016.

\bibitem{oleksenko18atc}
Oleksii Oleksenko, Bohdan Trach, Robert Krahn, Mark Silberstein, and Christof
  Fetzer.
\newblock Varys: Protecting {SGX} enclaves from practical side-channel attacks.
\newblock In {\em {USENIX} Annual Technical Conference ({ATC})}, pages
  227--240, 2018.

\bibitem{cosmix:atc19}
Meni Orenbach, Yan Michalevsky, Christof Fetzer, and Mark Silberstein.
\newblock Cosmix: A compiler-based system for secure memory instrumentation and
  execution in enclaves.
\newblock In {\em 2019 {USENIX} Annual Technical Conference ({USENIX} {ATC}
  19)}, pages 555--570, Renton, WA, July 2019. {USENIX} Association.

\bibitem{schwarz17dimva}
Michael Schwarz, Samuel Weiser, Daniel Gruss, Cl{\'{e}}mentine Maurice, and
  Stefan Mangard.
\newblock Malware guard extension: Using {SGX} to conceal cache attacks.
\newblock In {\em International Conference on Detection of Intrusions and
  Malware, and Vulnerability Assessment - 14th International Conference
  ({DIMVA})}, pages 3--24, 2017.

\bibitem{tsgx:ndss17}
Ming-Wei Shih, Sangho Lee, Taesoo Kim, and Marcus Peinado.
\newblock T-sgx: Eradicating controlled-channel attacks against enclave
  programs.
\newblock In {\em Network and Distributed System Security Symposium 2017
  (NDSS'17)}, February 2017.

\bibitem{shinde16asiaccs}
Shweta Shinde, Zheng~Leong Chua, Viswesh Narayanan, and Prateek Saxena.
\newblock Preventing page faults from telling your secrets.
\newblock In {\em {ACM} Asia Conference on Computer and Communications
  ({AsiaCCS})}, pages 317--328, 2016.

\bibitem{panoply:ndss17}
Shweta Shinde, Dat Le~Tien, Shruti Tople, and Prateek Saxena.
\newblock Panoply: Low-tcb linux applications with sgx enclaves.
\newblock In {\em NDSS}, 2017.

\bibitem{tsgxrepo}
{SSLab@Gatech}.
\newblock {T-SGX}.
\newblock \url{https://github.com/sslab-gatech/t-sgx}.
\newblock Accessed on 22/11/2021.

\bibitem{eddsa}
O~Sury and R~Edmonds.
\newblock Edwards-curve digital security algorithm (eddsa) for dnssec.
\newblock Technical report, RFC 8080 (Proposed Standard). Internet Engineering
  Task Force, 2017.

\bibitem{sgx-step}
Jo~Van~Bulck, Frank Piessens, and Raoul Strackx.
\newblock Sgx-step: A practical attack framework for precise enclave execution
  control.
\newblock In {\em Proceedings of the 2nd Workshop on System Software for
  Trusted Execution}, pages 1--6, 2017.

\bibitem{wang17ccs}
Wenhao Wang, Guoxing Chen, Xiaorui Pan, Yinqian Zhang, XiaoFeng Wang, Vincent
  Bindschaedler, Haixu Tang, and Carl~A. Gunter.
\newblock Leaky cauldron on the dark land: Understanding memory side-channel
  hazards in {SGX}.
\newblock In {\em {ACM} {SIGSAC} Conference on Computer and Communications
  Security ({CCS})}, pages 2421--2434, 2017.

\bibitem{sgx:eval}
Ofir Weisse, Valeria Bertacco, and Todd Austin.
\newblock Regaining lost cycles with hotcalls: A fast interface for sgx secure
  enclaves.
\newblock {\em ACM SIGARCH Computer Architecture News}, 45(2):81--93, 2017.

\bibitem{xu15sp}
Yuanzhong Xu, Weidong Cui, and Marcus Peinado.
\newblock Controlled-channel attacks: Deterministic side channels for untrusted
  operating systems.
\newblock In {\em {IEEE} Symposium on Security and Privacy ({SP})}, pages
  640--656, 2015.

\end{thebibliography}

\end{document}